\documentclass[referee,a4paper,12pt,traditabstract]{swsc}

\usepackage{url}
\usepackage[authoryear,round]{natbib}
\usepackage[xdvi]{graphicx} % substitude [dvipdf] or dvips above as needed
\newcommand{\Bl}{B_{\rm los}}
\newcommand{\Bz}{B_{\rm z}}
\newcommand{\Br}{B_{\rm r}}

\newcommand{\gradh}{\mbox{\boldmath$\nabla_h$}}
\newcommand{\nci}{{\tt NCI}}
\newcommand{\daffs}{{\tt DAFFS}}

\newcommand{\True}{{\rm H\&KSS}}
\newcommand{\TSS}{{\rm H\&KSS}}
\newcommand{\Appleman}{{\rm ApSS}}
\newcommand{\ApSS}{{\rm ApSS}}
\newcommand{\Brier}{{\rm BSS}}
\newcommand{\BSS}{{\rm BSS}}
\newcommand{\RC}{{\rm RC}} % rate correct
\newcommand{\CC}{C1.0+, 24\,hr}
\newcommand{\MM}{M1.0+, 24\,hr}
\newcommand{\XX}{X1.0+, 24\,hr}

\begin{document}

\title{The NWRA Classification Infrastructure: Description and Extension to the 
Discriminant Analysis Flare Forecasting System (DAFFS)}

\titlerunning{NWRA Classification, Solar Flares, and DAFFS}
\authorrunning{Leka, Barnes, and Wagner}

\author{K.D. Leka \inst{1} \and Graham Barnes \inst{1} \and Eric L. Wagner \inst{1}}
\institute{NorthWest Research Associates, Boulder Office \\
3380 Mitchell Ln. Boulder, CO 80301, USA. \\
\email{leka@nwra.com, graham@nwra.com, wagneric@nwra.com}}

\abstract
{A classification infrastructure built upon Discriminant Analysis has been
developed at NorthWest Research Associates for examining the statistical
differences between samples of two known populations.
Originating to examine the physical differences between flare-quiet and
flare-imminent solar active regions, we describe herein some details of the
infrastructure including: parametrization of large datasets, schemes
for handling ``null'' and ``bad'' data in multi-parameter analysis, 
application of non-parametric multi-dimensional Discriminant Analysis,
an extension through Bayes' theorem to probabilistic classification,
and methods invoked for evaluating classifier success.  The classifier
infrastructure is applicable to a wide range of scientific questions in 
solar physics. We 
demonstrate its application to the question of distinguishing flare-imminent
from flare-quiet solar active regions, updating results from the original
publications that were based on different data and much smaller sample sizes.
Finally, as a demonstration of ``Research to Operations'' efforts in the space-weather 
forecasting context, we present 
the Discriminant Analysis Flare Forecasting System (DAFFS), a near-real-time 
operationally-running solar flare forecasting tool that was developed from the 
research-directed infrastructure.}


\maketitle

\section{Introduction}
\label{sec:intro}

The prospect of forecasting rare events such as solar flares is a daunting
one, especially in situations where the exact trigger mechanism or
threshold for instability is not yet known.  Still, due to their impact
on the space weather environment, the forecasting of solar flares enjoys
some prominence of priority due to the combination of the speed-of-light
impact due to X-ray emission, their association with high-energy particle
enhancements, and their correspondence to coronal mass ejections.

Early empirical efforts to forecast solar flares focused on the 
white-light morphology of their host active regions \citep{flareprediction}.  Different classes
of active region (based on size and sunspot-group characteristics)
were observed to produce flares at different rates, and applying 
Poisson statistics resulted in probabilistic forecasts for 
flares \cite{McIntosh1990,BornmannShaw1994}.  This approach forms
the basis for many forecasts published today, including those from the 
US National Oceanic and Atmospheric Administraton / Space Weather Prediction 
Center \citep{gallagheretal02,Murray_etal_2017,flareprediction,RobChrisPC2017}

Empirical event-based forecasting is an application
of statistical classifiers to historical samples drawn from two known
populations: those that did and those that did not produce the event in
question.  
For statistical classification methods, there are essentially
four steps: (1) defining the event in question and hence the populations to be
sampled, (2) sample acquisition
(data acquisition), with attention to bias that may be imposed to the
samples of the populations, and parametrization of the data in such a
way as to be testable by the classifier, (3) applying the classifier
with appropriate safeguards against undue influence from outlier data
and statistical flukes, (4) evaluate the classification by way of
validation metrics or similar measures.  Finally, (5) the results are
available for scientific understanding or operational forecasting.

We describe herein a classifier infrastructure that has been developed at
NorthWest Research Associates (NWRA) based on NonParametric Discriminant Analysis
(``NPDA'').  Discriminant analysis is a tool that has been used for a variety of 
scientific investigations \citep{da_other_1,da_other_2,da_other_3}; it is 
particularly useful for the analysis of statistically-significant
samples of what are believed to be two distinct populations,
asking how well are the populations separable in parameter space?
While the NWRA Classification Infrastructure (``\nci'') has been
used for topics including detecting solar emerging flux regions
\citep{trt_emerge1,trt_emerge3}, and filament eruptions \citep{Barnes_etal_2017}, 
it developed from a series of works which examined the question, ``what is the
difference between a flare-imminent and flare-quiet solar active
region?'' \citep{dfa,dfa2,dfa3,Welsch_etal_2009,Komm_etal_2011a}.   

We describe in \S\ref{sec:nci} the \nci\ research-focused infrastructure,
with which new questions, new data, parameters, approaches, {\it etc.}
are developed and evaluated.  We discuss framing the question and event
definition considerations (\S\ref{sec:question}), data
analysis and parametrization (\S\ref{sec:data_params}), the various
flavors of Discriminant Analysis employed (\S\ref{sec:da}) and the
extension to probabilistic classification by way of Bayes' theorem
(\S\ref{sec:bayes}), evaluation tools (\S\ref{sec:evaluation}), and a
discussion of interpreting the results and selecting good parameter
combinations (\S\ref{sec:combos}).

In \S\ref{sec:nci_flares} we present a detailed example of \nci\ in use,
specifically NWRA's ongoing research regarding flare-imminent active
regions.  Following the steps above, this includes a description of the
event definitions (\S\ref{sec:nci_event_defs}), data sources and
parametrization modules (\S\ref{sec:nci_data}) including parametrization of 
temporal evolution (\S\ref{sec:dt_or_notdt}), classifier application examples 
(\S\ref{sec:nci_da}) and evaluation results (\S\ref{sec:nci_eval}).
We discuss the results of this updated flare-imminence research
in \S\ref{sec:nci_results}.

Flare forecasting research tools are becoming somewhat common
\citep{gallagheretal02,GeorgoulisRust2007,Wheatland2005,
BobraCouvidat2015,ColakQahwaji2009, Bloomfield_etal_2012}
and their performance an active subject of evaluation 
\citep[{\it{e.g.}}, ][]{BarnesLeka2008,Bloomfield_etal_2012,Falconer_etal_2014,allclear}.
As a demonstration of Research-to-Operation efforts in the space
weather forecasting context, we finally describe here the details
of the Discriminant Analysis Flare Forecasting System (``\daffs'',
\S\ref{sec:daffs}), a near-real-time operationally-designed forecasting
tool which grew from the \nci.  \daffs\ was recently implemented under a
Phase-II Small Business Innovative Research project through the National
Oceanic and Atmospheric Administration / Space Weather Prediction Center,
as an answer to ``Delivering a Solar Flare Forecast Model that Improves
Flare Forecast (Timing and Magnitude) Accuracy by 25\%.'' (topic NOAA
2013-1 9.4.3W).  It is presently in use to aid target selection for the
{\it Hinode} mission \citep{hinode}, specifically its limited field-of-view instruments
(the Solar Optical Telescope, \citet{hinode_sp}, and the EUV Imaging Spectrograph, \citet{hinode_eis}).

\section{The NWRA Classification Infrastructure \nci }
\label{sec:nci}

\begin{figure}
\centerline{\includegraphics[height=0.98\textwidth,clip, trim = 15mm 15mm 125mm 20mm, angle=-90]{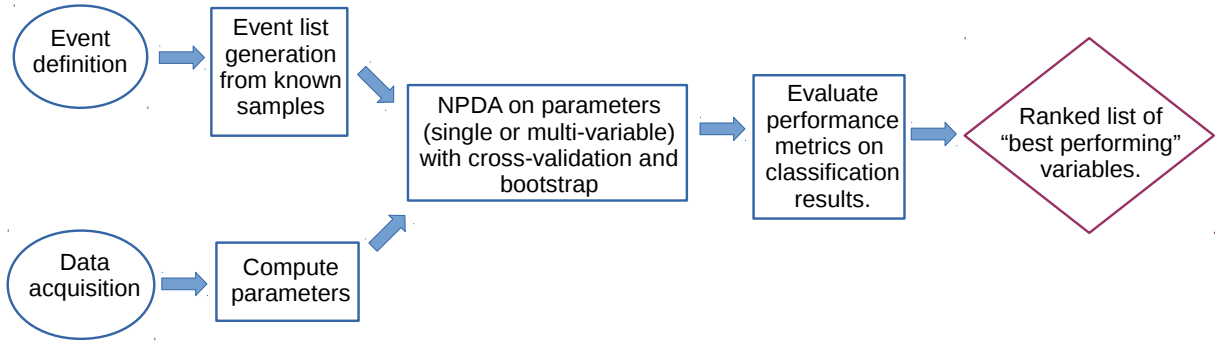}}
\caption{A very generalized flow chart for the NWRA Classification Infrastructure.  Circles generally 
indicate input, squares are processes, and the diamond is output.}
\label{fig:flow_nci}
\end{figure}

To understand a physical phenomena and guide relevant modeling efforts,
it is frequently important to identify what features are unique or
predisposed to an event.  \nci\ is a tool for doing that.  We describe
\nci\ here in general terms, guided by the steps outlined above (a general
flow-chart is provided in Figure~\ref{fig:flow_nci}).

\subsection{Posing the Question}
\label{sec:question}

A classifier attempts to separate samples from known populations in the
context of the parameter space constructed by variables which describe
some physical aspect of the system in question.  As such, the questions
posed must be constructed in such a way as to be addressable with a 
statistical classifier.  Classifiers can, for example, answer ``are these
three things uniquely associated with an event?'' but cannot answer, ``does 
this thing {\it cause} that event to occur?''  The crux of posing an 
appropriate question rests on the event definition.

\subsubsection{Event Definitions}
\label{sec:event_defs}

The event definition is simply the categorical description of an ``event'' and
the countering ``non-event''.  In solar physics, event definitions
have included whether (or not) a solar active region emerged
\citep{trt_emerge3}, or whether a filament produced a Coronal Mass
Ejection \citep{Barnes_etal_2017}.  A widespread application has
been regarding the occurrence (or not) of a solar flare within the context of
photospheric magnetic field measurements \citep{dfa,dfa3}, plasma velocity
\citep{Welsch_etal_2009} and helioseismology-derived parametrizations
\citep{Komm_etal_2011a}.

An Event Definition can be any such description, the more refined and specific
the better.  It must uniquely assign data points to one of the two populations
(event/non-event) against which the ability to distinguish the populations may
be judged.

By default, the \nci\ events are defined in a true forecasting sense,
with an upcoming interval during which the timing of an event is unknown.
However, \nci\ is designed for flexibility for scientific investigations.
With the construction of suitable event definitions and event lists,
\nci\ can also be invoked in an super-posed epoch analysis mode
\citep[``SPE''; ][]{trt_emerge1,trt_emerge3} for which the event time
is known and analysis is carried out relative to that reference time
\citep{MasonHoeksema2010,Reinard_etal_2010,BobraCouvidat2015}.

For scientific investigations the emphasis could be on understanding
the physical differences between populations. Such a study could use
either balanced sample sizes of the populations in question or invoke
equal prior probabilities for the two populations, in order to highlight
distinguishing characteristics \citep{trt_emerge1}.  This is in contrast
to a forecasting system where unequal sample sizes are the norm, and
one must incorporate the prior probabilities into the analysis.

\subsection{Data and Parametrization}
\label{sec:data_params}

Statistical classifiers attempt to separate samples drawn from different populations
within parameter space.  The samples are thus representations of the physical state
of the systems in question.  

For image-based or otherwise spatially distributed variables,
\nci\ derives both extensive and intensive
parameters, meaning those that do and do not depend on the size of the
feature in question, respectively \citep[see ][]{params,dfa,dfa3,Welsch_etal_2009}.
The spatially distributed variables ({\it i.e.} ``$x$'') are
parametrized by the first four moments: mean $\overline{x}$, standard
deviation $\sigma(x)$, skew $\varsigma(x)$, and kurtosis $\kappa(x)$, often weighted
by some relevant quantity, plus spatial summations as appropriate.  The
lower order moments capture the typical properties of the target image,
while the higher-order moments capture the presence of
small-scale features.
The image data themselves can be highly processed derived physical
quantities which have a 2-dimensional extent: in the case of imaging
spectroscopy, for example, an image of equivalent widths is appropriate.
For helioseismology, the image may be phase-shifts or inferred vorticity
maps \citep{Komm_etal_2011a,trt_emerge3}.

\nci\ is routinely used to analyze time-series data.  The approach
taken thus far is to fit time-series data with an appropriate model and
then use the retrieved coefficients or model fits as parametrizations.
In the case of the evolution of photospheric magnetic field over the
course of a few hours, for example, a first-order function is fit to the
image-derived parameters, then the slope and intercept at a designated
future time are considered as two input parameters.

The classifier employed here works best with continuous variables,
although it can handle discrete values.
As discussed below (\S~\ref{sec:da}), correlated parameters do not add useful
information.

Sample size matters.  Sufficient data allows for robust estimates of the 
classifier performance, as demonstrated by the resulting error bars on 
the validation metrics (see \S~\ref{sec:metrics}). Small sample sizes are
especially problematic for multi-variable analysis.  While each situation 
is different, samples significantly fewer than 100 data points in any one
population are challenging.  Sample sizes may be unbalanced.

\subsection{Discriminant Analysis}
\label{sec:da}

Discriminant Analysis \citep[DA;][]{ken83} is the classifier employed here: DA is a statistical 
approach to classify new measurements as belonging to one of two populations
based on characterizing the probability density functions (``PDF'') from known examples.
In brief, DA divides parameter space into two regions based on where the probability
density of one population exceeds the other.  Key to success is estimating 
the PDFs well. There are two general approaches to estimating PDFs: parametric, 
in which the functional form of the distribution is assumed, and nonparametric,
in which it is estimated directly from the data. Both options are available in
the \nci.

\subsubsection{Parametric Density Estimation}
\label{sec:da_linear}

When each variable is assumed to have a Gaussian distribution, with the same
variance for each population, the discriminant boundary is a linear function of
the variables.  This is quite a strong assumption, and one which is known to be
routinely violated for many solar physics relevant parametrizations; in
practice, for some topics, the results have been found to not depend strongly
on the assumption \citep{dfa3}, especially for rarer events.  

Linear DA has the advantage of being able to consider large numbers of
variables simultaneously even for small sample sizes.  If the variables are
uncorrelated and in standardized form, the magnitude of each variable's
coefficient in the discriminant function gives its relative predictive power
\citep[e.g.,][]{klecka80}. If variables are correlated then their predictive
power will be shared. In practice diminishing returns are usually found beyond
at most 10 parameters, likely due to correlations among variables in the solar
contexts tested thus far \citep{dfa,dfa3,Komm_etal_2011a}.  The disadvantage to
linear DA is that it assumes a functional form for the probability distribution
which may not be valid, and for most parameters considered here is likely
invalid.  Thus linear DA is best suited to problems for which only small
samples are available, and in which large numbers of variables are needed to
make accurate classifications.

\subsubsection{Nonparametric Density Estimation}
\label{sec:da_npda}

In nonparametric DA (NPDA), no assumption is made about the distribution of any
parameter.  Instead, the distributions are estimated directly from the data.
This negates the need for making assumptions, but requires large sample
sizes to accurately estimate the distributions, especially when considering
the shape of the tails of the distribution, and especially when considering
multiple variables simultaneously \citep[see Table 4.2 ][]{Silverman86}.
As such, NPDA can only reasonably be
used for combinations of small numbers of variables.  Within \nci, NPDA is
generally employed with at most 2-variable combinations.

The \nci\ NPDA code presently estimates the probability density using a kernel
method with the Epanechnikov kernel; a single smoothing parameter is set based
on its optimum value for a normal distribution \citep{Silverman86}.  This works
well for distributions which are not too far from a normal distribution, but
does not accurately represent long tails of distributions without very large
sample sizes.  Using a single smoothing parameter has a tendency to
undersmooth the tails of a distribution when the peak of the distribution is
appropriately smoothed, or to oversmooth the peak if the tails are
appropriately smoothed.  This is a particularly important effect when multiple
variables are considered, as the volume of space occupied by the tails (where
key parameter differences can lie) grows rapidly with the number of variables,
and thus the difference between the density of observations near the peak and
in the tails becomes even more pronounced.  To counteract this, the \nci\ works
with the logarithm of positive definite variables with a large skew (or the
logarithm of the absolute value of negative definite variables with a large
negative skew).  NPDA is thus best suited to problems for which large samples
are available, and for cases in which only a small numbers of variables are
needed to make accurate classifications.  With sufficiently large sample sizes,
even the tails of a distribution will be well captured by this approach.

Since it is impractical to apply NPDA to large numbers of variables
simultaneously, it is often helpful to consider the performance of all possible
permutations of small numbers of variables to determine which ones are best
able to classify the data (see \S\ref{sec:combos}).  Combining correlated
variables does not substantially improve the performance of NPDA over each
variable used alone. 

\subsubsection{Extension to Probabilistic Forecasts}
\label{sec:bayes}

To predict the probability of a data point belonging to a given population,
rather than a categorical (yes/no) classification,
Bayes' theorem is invoked \citep{SWJ}.   The probability that a measurement ${\bf x}$ belongs to
population $j$ is given by:  
\begin{equation} \label{eqn:prob} 
P_j({\bf x}) = {q_j f_j({\bf x}) \over q_1 f_1({\bf x}) + q_2 f_2({\bf x})}.
\end{equation}
where $q_j$ is the prior probability of belonging to population $j$, $f_j({\bf
x})$ is the probability density function for population $j$, and (for the flare
study) $j=2$ refers
to the flaring population, while $j=1$ refers to the flare-quiet population.
This expression is valid for any well behaved probability density function $f$.

\subsubsection{Missing Data}

In \nci\ missing data ({\it e.g.} due to outages) is differentiated from 
``well-measured null values'' such as the length of a strongly-sheared
magnetic polarity inversion line in a unipolar region (with no such line).
Neither of these categories of data are ignored or removed; instead,
they are assigned the climatology derived for all other datapoints with
the same missing-data assignment.  For multiple-parameter analysis,
the climatology is determined by those regions which have the same {\it
combination} of data flags (good, null, bad/missing).  In this manner,
as much information as is possible (for example if one parameter is good
but the other is missing) can be used in the classification.

\subsection{Evaluation} 
\label{sec:evaluation}

Evaluating the results of a classification exercise is a multi-faceted
task that includes not just the calculation of the validation metrics
but estimating their uncertainty, identifying and removing bias,
and accounting for statistical flukes.  Challenges to this and other
statistical methods include: small and/or unbalanced sample sizes,
bias in the samples, and undue influence from outlier data.  However,
with the use of appropriate metrics and error estimates, these challenges
are not insurmountable to achieving valuable understanding.

\subsubsection{Metrics}
\label{sec:metrics}

The success of the classification exercise is judged quantitatively
using metrics that are standard in classification and forecast validation
\citep{Woodcock1976,JolliffeStephenson2003,Bloomfield_etal_2012,allclear}.
We focus on the following: the Peirce skill score, also known as the
True Skill Statistic, or Hanssen and Kuipers discriminant (``\True'')
and the Appleman skill score (``\ApSS'') for categorical forecasts, and
the Brier Skill Score (``\Brier'') for probability forecasts. The Rate
Correct (``\RC'') is included for completeness.  The full derivations
and descriptions can be found in the cited references; of note here are
the critical functions of each: the \True\ measures the discrimination
between the Probability of Detection and the False Alarm Rate, the
\Appleman\ measures the skill against the climatological forecast, and
the \BSS\ evaluates the performance of probabilistic forecasts against
observed occurrence.  All are normalized such that 1.0 is the highest
possible score and 0.0 represents no separation / no skill relative
to the appropriate reference.

The categorical skill scores require a probability threshold ($P_{\rm
th}$) above which an event is classified (or forecast) to occur.
By default \nci\ uses $P_{\rm th} = 0.5$,
which effectively optimizes the \Appleman\ because the errors of both types 
are treated equally.
As discussed by \citet{Bloomfield_etal_2012,allclear} the optimal \TSS\
occurs when $P_{\rm th}\approx R$, $R$ being the event rate.  The Relative
Operating Characteristic (``ROC'') curve tracks performance of the \TSS\ as a
function of varying $P_{\rm th}$; the Gini coefficient $G1$ 
\citep[or ROC Skill Score;][]{JolliffeStephenson2003} then quantifies the
ROC curve: $G1=2*A - 1.0$ where $A$ is the area under the ROC curve, and again $G1 = 1.0$ 
denotes a perfect score.  Given the sensitivity
of the \TSS\ to this $P_{\rm th}$, which is to some level a function of
customer priority between False Alarms and Misses, we report the Gini
coefficient as a concise summary of the behavior of this metric.

The \Brier\ is itself a summary
of Reliability plots which graphically display systematic under- and
over-forecasting through the comparison of the predicted probabilities
to observed frequency \citep[see ][ for a discussion]{allclear}.

\subsubsection{Removing Bias}
\label{sec:xval}

To make unbiased estimates of the performance of the tested
parameters, the \nci\ system uses cross-validation \citep{hil66}.
In cross-validation, one data point is removed from the sample of
size $n$, and the remaining $n-1$ data points are used to make a 
probabilistic forecast about or to classify the
removed point.  This process is then repeated for all $n$ data points
and the elements of the standard classification table can be filled in
({\tt ``True Positive'',``False Positive'',``False Negative'' and ``True
Negative''}).  Essentially, this approach minimizes the influence of
outliers classifying themselves, and is a particularly important process
when working with small sample sizes.

\subsubsection{Estimating Skill Score Uncertainty}
\label{sec:errors}

A bootstrap method is used to account for random errors
\citep{EfronGong1983}, employing 100 draws by default.  The draws are
performed randomly across all data (meaning the sample sizes in each
draw are equal to the sample sizes of the two populations for the data),
with points selected randomly with replacement (each sampled point is not removed
from the drawn-upon sample, it may be drawn again).  The resulting skill
scores are computed for each draw, and the uncertainties are estimated
by the standard deviation of the draws' scores.

Small samples will generally result in large uncertainties.  For the
examples given in \S~\ref{sec:nci_eval}, with a sample size of events of
40 (for {\tt X+1}) the typical uncertainty in a skill score is 0.06 --
0.10, while with a sample of events of 2636 (for {\tt C+1}), the typical
uncertainy is 0.01. In both cases, the sample size of non-events is more
than a factor of 10 larger than the sample size of events, so the dominant
source of the uncertainty is the density estimates for the events.

\subsubsection{Accounting for Statistical Flukes}
\label{sec:flukes}

With a large number of parameters being tested, even with cross-validation
and bootstrap error estimates, statistical flukes can occur such that a
parameter appears to work well when it in fact does not.  The best remedy
for this is sample size, although the Sun does not always cooperate.

As described in \citet{trt_emerge3}, a Monte Carlo experiment can be used
to check that the outcome is not simply a result of random chance: two
random samples with sizes equal to the sample sizes of the two samples are
drawn from the same normal distribution to represent one variable with no
power to distinguish the two populations.  This is repeated for a large
number of variables, and the same analysis is performed on these random
variables as on the actual parameters being studied.  In this manner,
it can be determined what statistical outliers in the skill scores may
be expected given {\it no} difference in the two underlying populations
(see \S~\ref{sec:nci_eval}).

\subsection{What Was Learned?}
\label{sec:combos}

After the four steps above, the final one allows for analysis of what are
the physically important parameters for the physical question at hand,
{\it i.e.} ``what variables provide the most insight into the physical
processes at play in the context of the posed question?''  Alternatively,
in the operational context, determining the best performing parameters
is tantamount to producing a good forecast.

For many of the scientific questions addressed and addressable in \nci,
hundreds of parameters calculated from the data are considered, which
leads to millions of available parameter combinations.  In practice only
1- or 2-variable combinations are most appropriate for use, especially
with NPDA, where sample-size requirements increase very quickly with the
number of variables considered simultaneously.  However, with so many
parameters under consideration, statistical flukes will occur: variables or
variable combinations may seem to perform well (or poorly) when in fact
they do not, upon testing further with larger sample size.  As noted previously,
multi-parameter combinations formed from correlated variables do not add
discriminating power \citep{dfa3}.  

\subsubsection{Identifying Multiple Well-Performing Parameters}

To choose the best-performing single variable, NPDA is performed on
all available data including cross-validation and 100-draw bootstrap as outlined
above.  The single variables with the highest values of the skill score of interest
are selected for 1-variable results.  Bootstrap draws with no samples in one
or all populations (possible for small sample sizes) have a flag assigned 
for the skill scores, to either include in later computation or to ignore.

We developed a practical approach to evaluate multi-variable combinations
which likely captures the best-performing 2-variable combinations,
although it remains a possibility that a well-performing combination
is missed.   We first compute the skill scores for all possible
2-variable combinations without a bootstrap and using a simplified form
of cross-validation in which the contribution from the removed point
is subtracted from the density estimate, but the full density estimate
is not recomputed with an updated value of the smoothing parameter. Six
variables which appear in almost every combination of the top performing
combinations are selected.  Those are then paired with each of the others
available with the full cross-validation and bootstrap draws applied,
and the results are sorted on the skill score of choice.  There are
generally parameters or parameter combinations which perform similarly
within the uncertainties \citep{dfa}.  This is especially true for
classes having larger sample sizes.  The point of reporting multiple
parameter-combination (classification, forecast) results is not an 
ensemble result but a guard against statistical flukes (see below).

Different event definitions (\S\ref{sec:event_defs}) will lead to
different top-performing parameters or parameter combinations.  Similarly,
optimizing the results according to different skill scores will result
in different parameters being chosen \citep{allclear}.  In practice,
however, while there may be a slight re-ordering of the results, we
find that the same parameters (or very closely related parameters,
such as total magnetic flux {\it vs.} total negative magnetic flux)
are routinely within the top few percent of the results.

Having identified the best-performing parameters (or parameter combinations)
and their discriminating power one is then poised to physically interpret 
the results in the context of the posed question.  Conversely, if there are no results 
which indicate skill, then none of the parameters tested are able to successfully 
distinguish the populations in parameter space, and are essentially irrelevant
to the question posed.  Identifying the well-performing parameters also 
serves as a method for selecting those to be used for successful
operational forecasting.

\section{\nci\ and Empirical Research into the Causes of Solar Flares}
\label{sec:nci_flares}

As mentioned above, \nci\ originated from early research into the
statistical differences between flare-imminent and flare-quiet
time intervals \citep{dfa,dfa2,dfa3}.  The research has continued with new data
sources and parameters; as such, \nci\ has proven useful as a testbed infrastructure
for this research topic.  Here we demonstrate \nci\ with this specific
flare-related topic, and thus provide an update of the state of NWRA's research from
earlier publications.

\subsection{Event Definitions}
\label{sec:nci_event_defs}

\begin{figure}
\centerline{\includegraphics[width=0.50\textwidth,clip, trim = 0mm 0mm 0mm 0mm, angle=0]{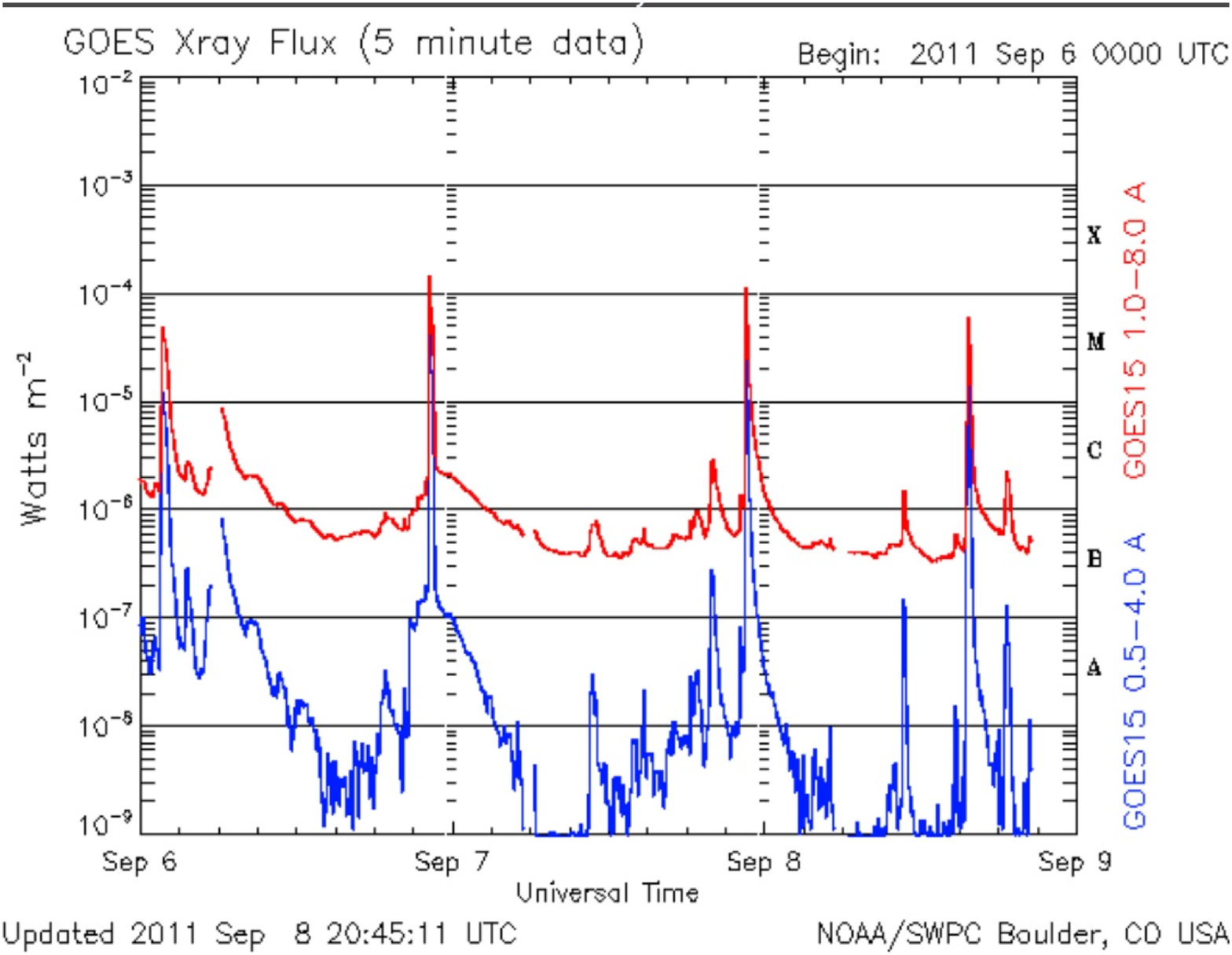}
\includegraphics[width=0.50\textwidth,clip, trim = 0mm 0mm 1mm 0mm, angle=0]{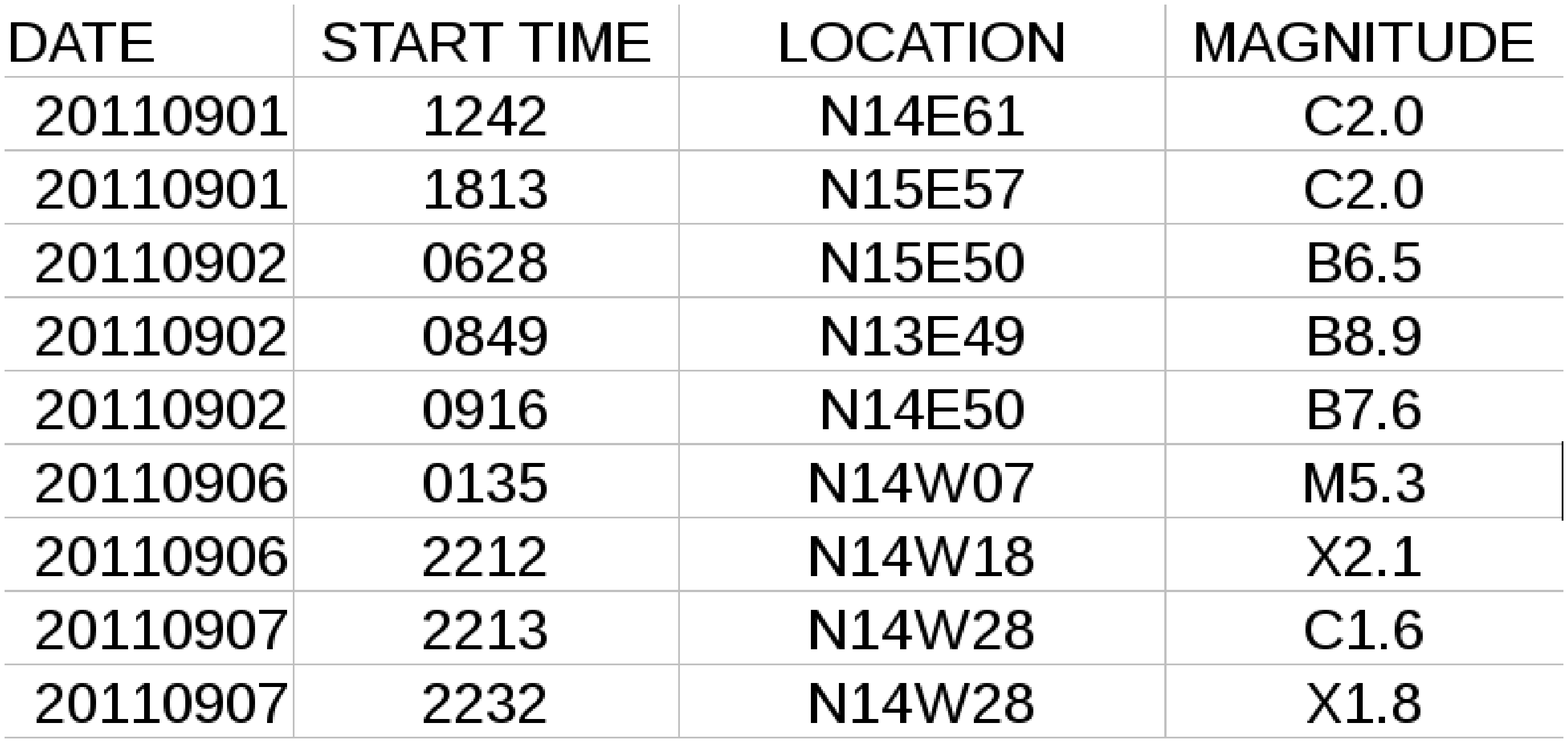}}
\caption{{\bf Left:} The GOES soft X-Ray plots of the 2011 Sep.~6--9 interval.
{\bf Right:} The event list for flares detected in the GOES soft X-Ray
1--8\AA\ channel, specifically for NOAA~AR~11283, through 20110907
(Active Region and time interval chosen fairly randomly, for additional demonstrations below).}
\label{fig:11283events}
\end{figure}

For the \nci\ research on flare-imminent active regions, we adopted a 
``forecast'' model for event definitions, and 
essentially match the forecasts issued by NOAA, as summarized
in Table~\ref{tbl:eventdef} based on events as illustrated in 
Figure~\ref{fig:11283events}.  

\begin{table}
\caption{Event Definitions}
\begin{center}
\begin{tabular}{l c c c c c }
Label	    & minimum  & minimum & latency & validity & Region/ \\
	& threshold   & threshold & & period & Full Disk \\
	& (nomenclature)   & (${\rm Watt\,m}^{-2}$) & (hr) & (hr) & \\ \hline
{\tt C+1} & C1.0 & $\ge 10^{-6}$ & 2.2 & 24 & R, FD{$^\dagger$} \\
{\tt M+1} & M1.0 & $\ge 10^{-5}$ & 2.2 & 24 & R, FD  \\
{\tt X+1} & X1.0 & $\ge 10^{-4}$ & 2.2 & 24 & R, FD  \\ \hline
{\tt C+2}{$^\dagger$} & C1.0 & $\ge 10^{-6}$ & 26.2 & 24 & R, FD  \\
{\tt M+2} & M1.0 & $\ge 10^{-5}$ & 26.2 & 24 & R{$^\dagger$}, FD  \\
{\tt X+2} & X1.0 & $\ge 10^{-4}$ & 26.2 & 24 & R{$^\dagger$}, FD  \\ \hline
{\tt C+3}{$^\dagger$} & C1.0 & $\ge 10^{-6}$ & 50.2 & 24 & R, FD  \\
{\tt M+3} & M1.0 & $\ge 10^{-5}$ & 50.2 & 24 & R{$^\dagger$}, FD  \\
{\tt X+3} & X1.0 & $\ge 10^{-4}$ & 50.2 & 24 & R{$^\dagger$}, FD  \\ \hline
\end{tabular}
\flushleft{$\dagger$:not included in NOAA's forecasts but included here for completeness.}
\end{center}
\label{tbl:eventdef}
\end{table}

The thresholds are defined by the peak Soft X-ray flux as reported by
the National Oceanic and Atmospheric Administration (NOAA) Space Weather
Prediction Center (SWPC), based on the 1--8\AA\ Soft X-ray detector on the
{\it Geostationary Observing Earth Satellite} (``GOES'') platform at the
time of the event.  It should be noted that in the application of DA to
flaring active regions as defined in this manner, a continuous variable
(the Soft X-ray emission) defines the categorical event definition
required by DA by means of a fairly arbitrary threshold.  Additionally,
defining an event by its peak Soft X-ray emission captures only a limited
measure of flare-associated energetic output.

We included a 2.2hr delay needed for all data processing (including
initial SDO/HMI data-reduction at Stanford) between data acquisition
time and forecast issuance time, although for research based on the
definitive time series, this is effectively moot.  In other words, when
relying upon the definitive data, the last dataset used is from {\tt
21:48 TAI} for a forecast (classification) issued at {\tt 00:00 UT}.
The SXR events considered to match the forecast validity periods, are
taken from that issuance time.

The region-by-region forecasts are based on the 
HMI Active Region Patches \citep[`HARPs'][ see \S\ref{sec:hmi}]{hmi_pipe,hmi_sharps} 
rather than
strictly on NOAA Active Regions (ARs).  This approach has two consequences
if the results are compared to strictly NOAA AR-based methods: first,
there are many HARPs which contain more than one NOAA region.  As such,
the distributions of parameters (in particular extensive parameters such
as total magnetic flux) will have an extended large-value tail compared
to single ARs.  Second, most HARP numbers are assigned to regions which
do not have a corresponding NOAA number.  Such regions are plentiful,
are predominantly small (often consisting solely of plage), and hence the
parameters (both extensive and intensive) will have distributions which
are more densely populated at lower values than AR-based distributions.

The full-disk forecasts are created by combining the flaring probabilities
of the regions:

\begin{equation}
P_{\rm FD} = 1.0 - \prod_{\rm AR} (1.0 - P_{\rm AR})
\label{eqn:fdar}
\end{equation}

\noindent
where $ P_{\rm AR}$ are the flaring probabilities of each active region 
and $P_{\rm FD}$ is the resulting full-disk flaring probability.
Of note, combining probabilities in this manner assumes independent
events; this assumption may not hold during times of high solar activity
when numerous magnetically interconnected regions are present on the
solar disk \citep{SchrijverHiggins2015}.  The \nci\ can thus provide full-disk forecasts but those
are usually less useful for research purposes and more appropriate for
the operational tool (see \S~\ref{sec:daffs}).

\subsection{Data Sources and Parametrization}
\label{sec:nci_data}

Early NWRA \nci-based research examined the question of
differences in flare-imminent {\it vs.} flare-quiet active regions
\citep{params,dfa,dfa2,dfa3} using a substantial database of time-series
photospheric vector magnetic field data from the U. Hawai`i Imaging
Vector Magnetograph \citep{ivm1,ivm2,ivm3,ivm4}.  Additional studies
\citep{allclear} used line-of-sight photospheric magnetic field data from
the Solar and Heliospheric Observatory / Michelson Doppler Imager \citep[SoHO/MDI][]{mdi}.  
Current NWRA research and performance baselines
use new data sources, as described here.

To characterize the state of the active regions
and their likelihood to flare, we perform parametric analysis on the
data sets described above.  The goal is to reduce the ($[x,~y,~t]$)
vector magnetic field time series plus the flaring history of each active
region into a series of parameters suitable for statistical analysis.
The parametrization packages which are now well-established or are
under active research are fairly modular; the results from each are
often merged after the parameters are calculated, for classifier analysis.

\subsubsection{Data: NOAA-Generated Soft X-Ray Event Lists}
\label{sec:event_list}

NWRA gathers the event lists of solar flares as defined by NOAA/SWPC;
the information required includes the start time, peak time, peak flux
and associated NOAA-numbered active region source (Figure~\ref{fig:11283events}).  
These are acquired
from the NOAA archives through {\tt ftp.swpc.noaa.gov/pub/warehouse}
and from the near-real-time updating lists at 
{\tt http://services.swpc.noaa.gov/text/solar-geophysical-event-reports.txt}.
Additionally, NWRA subscribes to the NOAA {\it External Space Weather
Data Store} (``E-SWDS'') near-real-time space weather data access system,
where events are recorded with minimal delay.  We do not attempt to
perform any of the analysis by which to independently determine start
time, peak flux, or location.  In particular, flares for which NOAA
does not provide an associated active region, as occurs for a fraction
of smaller flares and a few larger flares, are not included in the \nci\ analysis by default.

\subsubsection{Data: SDO/HMI Photospheric Vector Magnetic Field Data}
\label{sec:hmi}

The present \nci\ research extends earlier work naturally by
relying upon time-series of vector magnetic field data from the
Helioseismic and Magnetic Imager from the Solar Dynamics Observatory
\citep{sdo,hmi,hmi_pipe,hmi_invert,hmi_cal}.
NWRA became an SDO remote SUMS/DRMS (Storage Unit Management System/Data
Record Management System) site early in the mission, as NWRA's
``ME0'' disambiguation code is used for the pipeline data reduction
\citep{hinode2ambig,hmi_pipe} and easy access to the data was needed for
pipeline-code implementation.  
NWRA maintains local copy of the definitive hourly HMI data through the 
remote-SUMS system and automated updates of the metadata DRMS database.
Definitive data are transferred roughly monthly using the JSOC Mirroring
Daemon (JMD\footnote{see \url{http://vso.tuc.noao.edu/VSO/w/index.php/Main_Page}}).

\begin{figure}
\centerline{\includegraphics[width=0.95\textwidth,clip, trim = 10mm 15mm 35mm 05mm, angle=0]{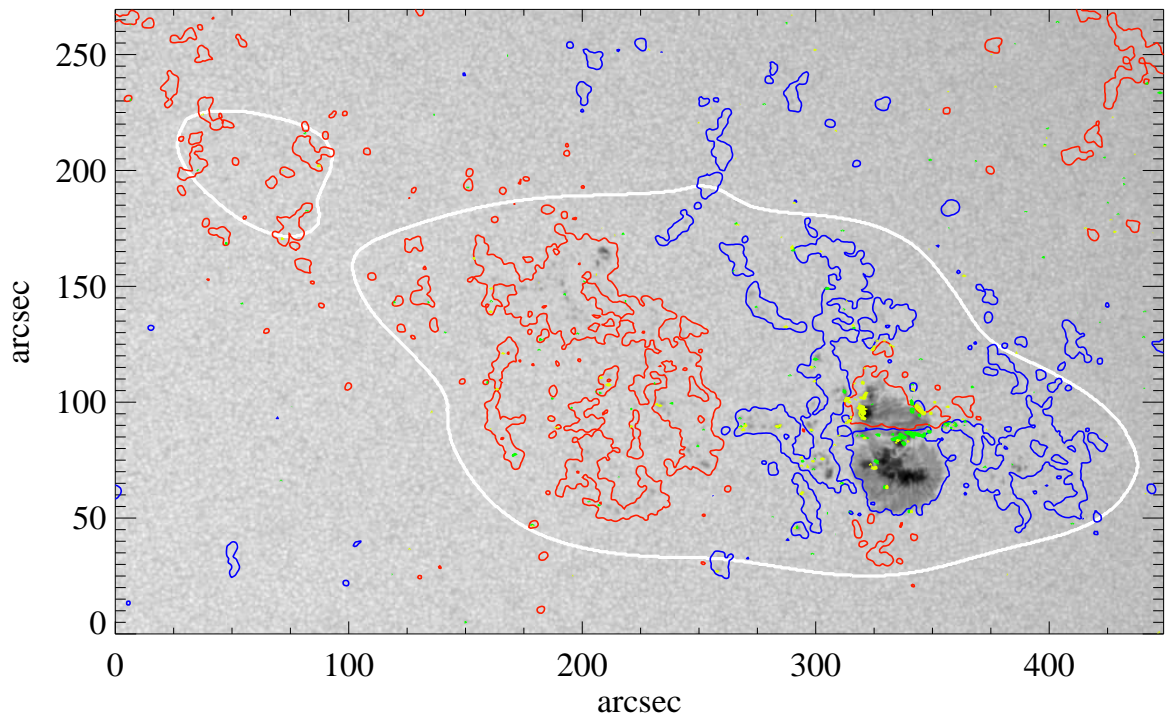}}
\caption{An image of NOAA~AR\,11283 at 2011.09.07 21:48\,TAI as
an {\it SDO}/HMI Active Region Patch (HARP \#833), showing the
vertical component of the magnetic field $B_z$ from {\it SDO}/HMI
(red/positive, blue/negative contours at $\pm 100\, {\rm Mx\,cm}^{-2}$); axes
are in arcsec, and green/yellow contours indicate the vertical current density
$J_z$ at $\pm 50,100\, {\rm mA\,m}^{-2}$.  
The HARP ``active pixel'' area is outlined in grey. }
\label{fig:11283gram}
\end{figure}

For the flare-targeted research with \nci\
the definitive data series are used; we do not generally employ the
cylindrical equal-area reprojected data.  The {\tt hmi.Mharp\_720s}
series, including the bitmap which flags active-pixel ``blobs''
\citep[see Figure 3 and Table 8; ][]{hmi_pipe} is used to identify the HARPs that
are extracted from the full-disk Milne-Eddington inversion data series
{\tt \tt hmi.ME\_720s\_fd10}, including uncertainties from the inversion
\citep{hmi_invert}.  The {\tt hmi.B\_720s} series provides the
pipeline disambiguation results and their confidence \citep{hmi_pipe}.
Additionally, we optionally extract the same HARP sub-areas from the
continuum images in {\tt hmi.Ic\_720s} and the line-of-sight magnetograms
in {\tt hmi.M\_720s}.  An example HARP and active-pixel map is shown
in Figure~\ref{fig:11283gram} for the region associated with the flares
demonstrated in Figure~\ref{fig:11283events}.

To keep the data volume and analysis tenable, an hourly cadence is
generally used; to avoid problems from daily calibrations that are
scheduled at some of the {\tt :00} times, we focus on the {\tt :48} data.
No data are used that have a quality flag different from zero.
As described in Section~\ref{sec:dt_or_notdt}, data covering
6hr inclusive (7 time periods) provide input for short-term
temporal-evolution parameters.  By default the \nci\ classification
mimics a midnight-issued forecast, thus hourly data from 15:48 -- 21:48
are targeted.  For definitive data, there is no constraint regarding
the HMI data reduction pipeline timing requirements.  

The hourly HARP-based sub-areas are extracted, coaligned, and gathered together
as custom in-house FITS files.  The filename describes the start time, length of 
time-series, cadence, and HARP number, and is unique; as such it forms the basis
for cross referencing throughout \nci.  The {\tt hmi.Bharp\_720s} and 
{\tt hmi.sharp\_720s} series (both linked to {\tt hmi.MEharp\_720s}) 
provide disambiguated extracted HARP-area vector field data cut-outs.

Following earlier studies \citep{params,dfa,dfa3}, the image-plane
coordinate system is used by default, with transformations to heliographic
components performed using point-by-point observing angle component
computation and a propagation of uncertainties based on the uncertainties
from the inversion as included in the {\tt hmi.ME\_720s\_fd10} series.
The multi-point approach avoids significant errors arising from a
planar assumption in the (not infrequent) case that a HARP extends over
a significant portion of a visible half-hemisphere.  A potential field
whose normal component matches the input normal component boundary is
computed using a Cartesian Fourier-component method; a guard ring which is a
factor of 2 larger than the input boundary size is invoked in order to
reduce artifacts which can arise with a boundary occurring on, or too
near, strong-field areas.  An estimate of the 1-$\sigma$ uncertainty in
the potential-field components is computed using the computed components
themselves: within active pixels, those that are brighter than the
mean less the mean absolute deviation of the continuum intensity
are used to compute histograms of that component's field strengths.
The field strength where 68\% of the area under the histogram curve
is achieved is denoted that component's 1-$\sigma$ uncertainty level.
The analysis of the vector field data extends to $\theta \approx 80^\circ$
from disk center.

\subsubsection{Data: Photospheric Line-of-Sight Magnetic Field Data}
\label{sec:blos}

Photospheric line-of-sight magnetic field data is used in some \nci\
research on solar flare productivity.  The SoHO/MDI mission provided
a significant data set, used in one of the first systematic forecasting
comparison efforts \citep{allclear}.
Presently, the Global Oscillations Network Group
\citep[``GONG''; ][]{gong}, in particular the line-of-sight magnetograms,
is an ingested dataset for analysis by \nci\ tools.
When fully implemented, the GONG data will be used in the Near-Real-Time
forecasting system in the event that the vector field data are
unavailable (see Section~\ref{sec:nrt_data}).  The GONG data do have
the advantage of lengthy coverage, which increases the
statistical significance of the training results. 

Data from all available GONG stations are acquired for a target time
from the NSO near-real-time GONG website, and assessed.  The best is
chosen based on metrics (moments of spatial gradients) which quantify
the seeing quality with a normalization to account for slightly differing
plate scales between the sites.

GONG-based active-region patches (``GARPs'') are extracted according
to the NOAA E-SWDS Active-Region data, with a bounding-box generally
reflecting the size of the region (latitude, longitude, and area),
then adjusted for solar rotation to the target time of the acquired data.

In general when $\Bl$ is employed for analysis, an estimate of $\Bz$ is derived 
using a potential field which matches the observed $\Bl$, following \citet{bbpot}. 
Note that this differs from assuming that $\Bl~\approx\Bz$.  From this potential field,
a radial field estimation, $\Bz^{\rm pot}$, is retrieved to be used for the input.
The advantage of this treatment of a $\Bl$ boundary is twofold, as discussed
in \citet{bbpot}: the magnetic field strengths better approximate the radial 
field strengths and the apparent magnetic polarity inversion lines that are in fact 
caused only by viewing angle are removed or significantly mitigated.
As such, for line-of-sight sources as with vector data sources, the magnetic field analysis
functions fully to $\theta \approx 80^\circ$.

\subsubsection{Other data sources}

For research purposes, the \nci\ has been used for investigations using 
helioseismology and coronal emission as related to emerging magnetic flux 
\citep{trt_emerge1,trt_emerge3} as well as flaring.  The data sources
for these investigations have been or will be described in detail in 
the relevant publications.

\subsubsection{Parametrization: Prior Flare History}
\label{sec:pff}

Future solar energetic events often follow previous activity,
with a somewhat predictable distribution in size and
frequency \citep{Wheatland2005}.  As such it is not surprising
that ``flare persistence'' can be a good indicator of future
activity \citep{Falconer_etal_2012}; in fact, it forms one of
the primary predictors used for NOAA/SWPC operational forecasts
\citep{flareprediction}.

From this module, the parametrization of prior flaring activity
provides event probabilities based on peak Soft X-Ray output over the
prior 6, 12, and 24\,hr, including the change in flaring output between
those intervals.  This ``prior peak flux'' (``PFF'') parametrization essentially
follows \citet{Abramenko2005b}:
\begin{equation} 
{\rm FL}=\sum {\rm index}_C+10\times{\rm index}_M+100\times{\rm index}_X 
\label{eqn:flareflux}
\end{equation}
\noindent
where ${\rm index}_C \equiv 2.3, 4.9$, {\it etc.}, as reported by the
NOAA GOES 1--8\AA\ detectors.  Example categories include:

\begin{itemize}
\item The total peak flux of the target active region in the 12~hr prior 
to the forecast, 
\item The total peak flux from the time the region was first identified 
prior to the forecast,
\item The number of flares produced by a target active region in the 
prior 12~hr compared to the prior 24~hr.
\end{itemize}

There is no additional statistical sophistication in these
parametrizations of flaring history.  While the comparisons and summations
are performed for the prior 6, 12, and 24\,hr intervals, there has not yet 
been any systematic testing of whether these terrestrially-defined periods
are optimal \citep{EEG_chapter}.

\subsubsection{Parametrization: Photospheric Magnetic Field}
\label{sec:nci_mag}

The parametrization of the photospheric magnetic field essentially follows
\citet{params,dfa,SWJ}.  The goal is to identify signatures of magnetic
field complexity, to estimate energy storage, and identify non-potential
structures and magnetic twist -- in the context that regions showing
evidence of minimal available magnetic energy, minimal energy storage,
low complexity and nearly-potential (current-free) fields are much less
likely to flare, and {\it vice versa}.  As described in \citet{SWJ},
the photospheric parameters in many ways quantify the characteristics
which are used for the sunspot-group classifications such as the McIntosh
\citep{McIntosh1990} (modified Zurich) nomenclature.  The variables,
described in detail in \citet{params}, fall into nine broad categories:
\begin{itemize}
\item the magnetic field vector component magnitudes, $B_z$ and $B_h$
\item the inclination angle of the fields, $\gamma\!=\!\tan^{-1}(B_z/B_h)$
\item the horizontal spatial gradients of the magnetic fields,
$|\mbox{\boldmath$\nabla$}\!_h B|$, $|\mbox{\boldmath$\nabla$}\!_h B_z|$, 
$|\mbox{\boldmath$\nabla$}\!_h B_h|$
\item the vertical current density, $J_z\!\sim\! (\partial B_y/\partial x -
\partial B_x/\partial y)$
\item the force-free parameter, $\alpha\!\sim\!J_z/B_z$
\item the vertical portion of the current helicity density, $h_c\!\sim\!J_z B_z$
\item the shear angle from potential, $\Psi\!=\!\cos^{-1} ({\bf B}^p {\mathbf{\cdot}} 
{\bf B}^o/B^p B^o)$
\item the photospheric excess magnetic energy density,
$\rho_e\!=\!({\bf B}^p - {\bf B}^o)^2/8\pi$
\item the magnetic flux near strong-gradient magnetic neutral lines, $\mathcal{R_{\rm nwra}}$
\end{itemize}
\noindent
where ${\bf B}^p$ is the computed potential field referred to in \S~\ref{sec:hmi}.
All quantities are based on physically-meaningful helioplanar components that
include or are weighted by the deprojected pixel area, as approrpiate.

The last category follows \citet{Schrijver2007}, who developed a parameter
to characterize current-carrying emerging flux systems 
which manifest as magnetic neutral lines displaying strong
spatial gradients:  
\begin{equation}
\mathcal{R_{\rm nwra}} = \sum f \ast \mathcal{B}(\gradh \Bz\pm)\, dA
\end{equation}
where $\mathcal{B}(\gradh \Bz\pm)$ identifies polarity-inversion lines
and $f$ is a Gaussian convolution function.  \citet{Schrijver2007} identified
magnetic neutral lines using polarity-specific bitmaps constructed
using {\it SoHO}/MDI \citep{soho,mdi} $\Bl$ magnetograms, including only
areas that exceeded a fixed $150{\rm Mx\, cm}^{-2}$ threshold, and dilating
the bitmaps with a $6^{\prime\prime} \times 6^{\prime\prime}$ kernel; overlapping
regions were thus identified as strong-gradient magnetic neutral lines.
A new bitmap with the neutral lines thus identified was convolved with a
Gaussian of FWHM $D_{\rm sep} = 15$Mm (fixed at 10 MDI pixels) to obtain a
weighting map which, when multiplied by the $|\Bl|$ magnetogram and summed
(assuming a fixed areal coverage of $2.2\times10^{16}{\rm cm}^2$ per
pixel), resulted in $\mathcal{R}$.  In \citet{Schrijver2007}, regions
were only considered within $45^\circ$ of disk center, with the assumption
$\Bl~\approx \Br$, and viewing-angle impacts on pixel-sampled area, etc,
were negligible \citep[see also ][]{allclear}.  Only strong-flare regions
were considered.

\begin{figure}
\centerline{\includegraphics[width=0.95\textwidth,clip, trim = 10mm 15mm 35mm 05mm, angle=0]{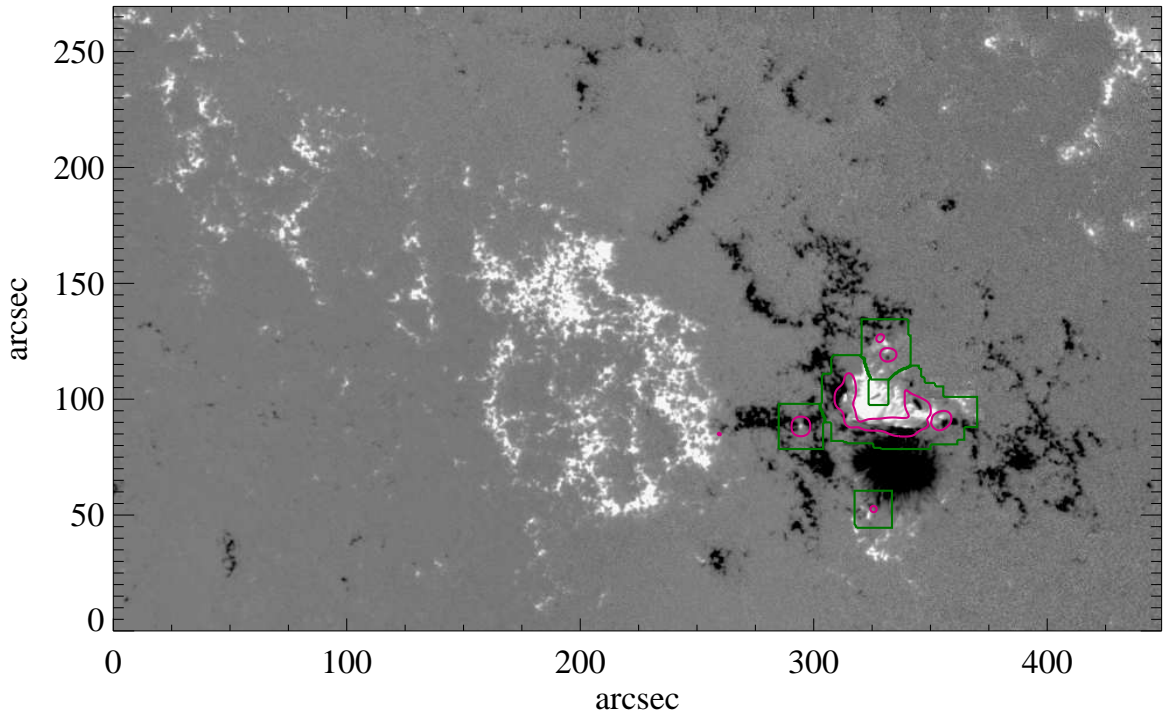}}
\caption{The vertical magnetic field strength (white/black being positive/negative, 
scaled to $500$G) of HARP\,833 {\it i.e.} NOAA~AR\,11283 ({\it c.f.} Figure~\ref{fig:11283gram}), 
showing regions identified as strong-gradient magnetic neutral lines for which
the $\mathcal{R_{\rm nwra}}$ parameter is calculated (green boxes) with the 0.68-level
contour of the Gaussian-convolved bitmap indicated (purple contours).  
Similar masks are used for the ``magnetic neutral line''-related parametrizations.}
\label{fig:pil}
\end{figure}

The approach taken here is to calculate a quantity $\mathcal{R_{\rm nwra}}$ 
which more closely represents the physically relevant boundary
properties: as such, the vertical component of the field $\Bz$ is used
rather than $\Bl$; the threshold is based on the noise in the $\Bz$
component (and thus varies), and all steps which depend on distances
or sizes use an appropriate physical distance in $10^6$\,m as computed
according to pixel size and observing angle (and thus the number of pixels
used will vary with observing angle).  Of note, there is no attempt
to identify only a single or primary strong-gradient neutral line, and
depending on sizes and thresholds chosen, numerous small strong-gradient
neutral lines may be identified (Figure~\ref{fig:pil}).  The resulting
magnitudes of $\mathcal{R_{\rm nwra}}$ differ from \citet{Schrijver2007},
in part due to the new implementation but also due to the different
data source, although the general behavior is very similar \citep[see
discussion in ][]{allclear}.

For each quantity, the parametrization only includes ``active pixels''
with the full HARP box.
For all magnetic field data, we include only those pixels with
signal/noise ratio $S/N > 4$ by default (the $S/N$ threshold is a
definable parameter).  Masks are created using these criteria, but
which are then eroded and dilated using a $2\times2$ shape operator in order
to avoid single-pixel inclusions.  For quantities that, for example,
require spatial derivatives, the derivatives are taken on the full
sub-area, then the appropriate computation (moment analysis, totals,
{\it etc.}) include only those pixels that meet the selection criteria.
Separate but analogous masks are created to identify the magnetic neutral
lines, using similar thresholds and boundary-smoothing approaches.

As described above, moment analysis and extensive parameters are used to
describe the spatially distributed variables.  The magnetic character
of an active region is thus reduced to approximately 150 variables.
In the event that the $\Bl$ data are used, only parameters that do not rely
on the horizontal component of the field are calculated.

\subsubsection{Parametrization: Magnetic Charge Topology}
\label{sec:mct}

\begin{figure}[t!]
\centerline{
\includegraphics[width=0.85\textwidth,clip, trim = 05mm 0mm 05mm 05mm, angle=0]{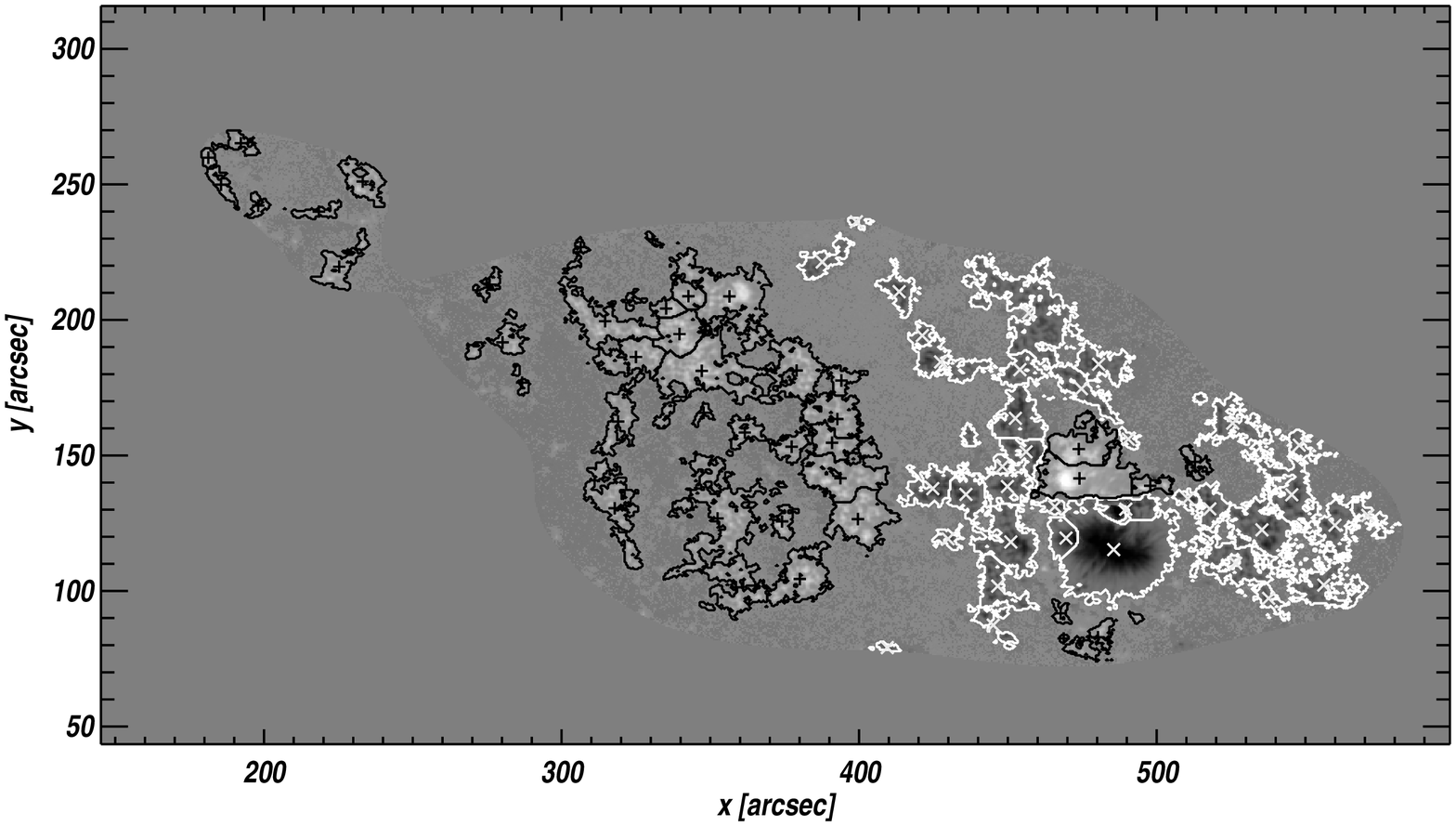}}
\centerline{
\includegraphics[width=0.85\textwidth,clip, trim = 05mm 0mm 05mm 05mm, angle=0]{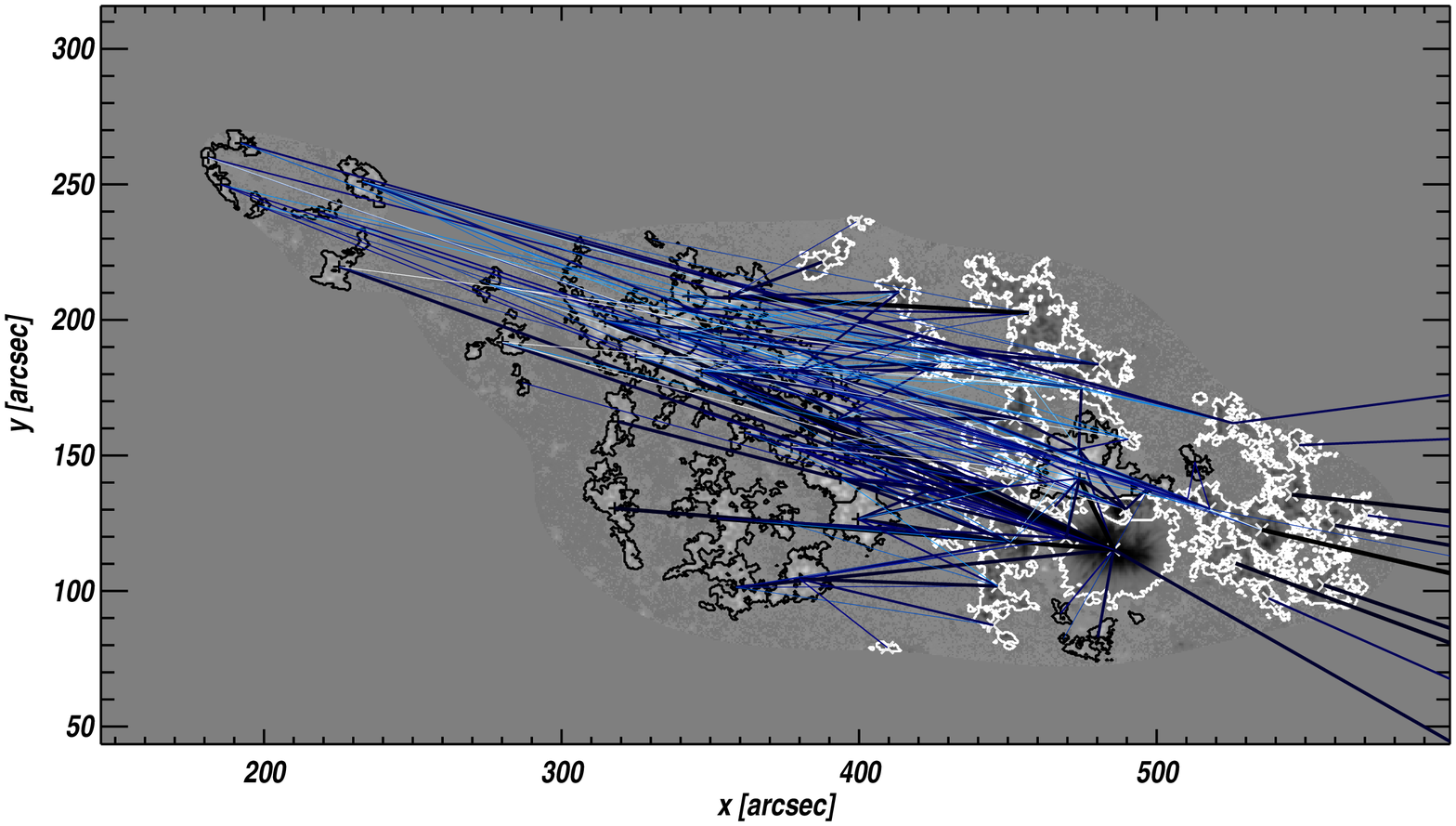}}
\vspace{-0.25cm}
\caption{Demonstration of Magnetic Charge Topology analysis,
characterizing the coronal magnetic topology of
NOAA~AR\,11283 (Figure~\ref{fig:11283gram}), 2011.09.07 at 21:48:00\,TAI.
{\bf Top:} Magnetogram showing the vertical field (greyscale),
the partitions (contours), and the locations of the sources ($+$ and
$\times$ for positive and negative).  Areas outside of the active-pixel
bitmap ({\it c.f.} Figure~\ref{fig:11283gram}) are set to zero.  {\bf Bottom:} same as top panel,
but schematically showing the connectivity matrix (blue lines) with
dark/thick lines indicating more flux in the connection than light/thin
lines.}
\label{fig:mct}
\hrulefill
\vspace{-0.5cm}
\end{figure}

Most solar energetic events are believed to ultimately originate in
the corona, where the magnetic field generally dominates the plasma
and the climate is more conducive to the storage and subsequent
rapid release of energy via magnetic reconnection.  A corona with a
very complex magnetic topology is one which should more readily allow
magnetic reconnection to initiate, and hence an eruptive event to begin.
The Magnetic Charge Toplogy model describes the coronal topology and its
evolution, using as a boundary time-series maps of the photospheric
radial field $\Bz$.  Concentrations of magnetic flux in an active
region are represented by point sources (Fig.~\ref{fig:mct}, top).
A gradient-based tessellation scheme, supplemented by the partitioning
of a reference time-averaged magnetogram, is used to track each magnetic
concentration with time \citep{mct}.  The coronal magnetic field is
modeled as the potential field of the point sources, from which a unique
magnetic connectivity matrix is derived (Fig.~\ref{fig:mct}, bottom).
While arguably inappropriate for very complex active regions, using a
potential field is fast (significantly faster than non-linear force-free
extrapolations), physics-based (see discussion in \citet{BarnesLeka2008}),
and arguably captures the key features of coronal complexity
associated with event-productive solar active regions \citep{Regnier2012}.

The MCT variables used in the \nci\ analysis \citep[see][]{dfa2}
focus on categories describing the distribution of:
\begin{itemize}
\item the number and separation of magnetic sources $S$
\item the flux assigned to each source, $|\Phi_i|$
\item the magnetostatic energy $E_B = {1 \over (2 \pi)^2} \sum_{i<j} {\Phi_i \Phi_j \over 
\vert {\bf x}_i - {\bf x}_j \vert}$
\item the characteristics of the connectivity matrix $\psi_{ij}$ describing the flux connecting source
$i$ to source $j$.
\item the flux in each connection weighted by inverse distance between
connected sources,
$\varphi_{ij} = {\psi_{ij} \over \vert {\bf x}_i - {\bf x}_j \vert}$
\item the angle between the north/south axis and the line segment between
connected sources,
$\xi_{ij} = \tan^{-1} \bigg [{x_j - x_i \over y_j - y_i} \bigg ]$ for
$\psi_{ij} \ne 0$.
\item the distribution of the number of connections from each source $C_i$
\end{itemize}
\vspace{0.1cm}
\noindent
The basic analysis calculates almost 50 parameters based on
the above-mentioned characteristics.
Of note, due to computational requirements the magnetic null-finding (as was 
described in \citet{mct}) is not routinely performed for large datasets.

As discussed in \citet{BarnesLeka2008}, $\varphi_{ij}$ is essentially
indistinguishable from the ``effective connected magnetic field'' 
$B_{\rm eff}={\psi_{ij} / \vert {\bf x}_i - {\bf x}_j \vert^2}$ 
\citep{GeorgoulisRust2007} with the primary distinguishing features being 
the square of the distance and more importantly, a physics-based potential field
forming the basis of the connectivity matrix for the MCT parameters. A 
quantity $\varphi_2$, which uses the potential-field based connectivity 
in the expression for $B_{\rm eff}$, is included in the \nci. 

\subsubsection{Static {\it vs.} Temporal Variation}
\label{sec:dt_or_notdt}

\begin{figure}
\begin{center}
\includegraphics[width=0.6\textwidth,clip, trim = 5mm 5mm 0mm 0mm, angle=0]{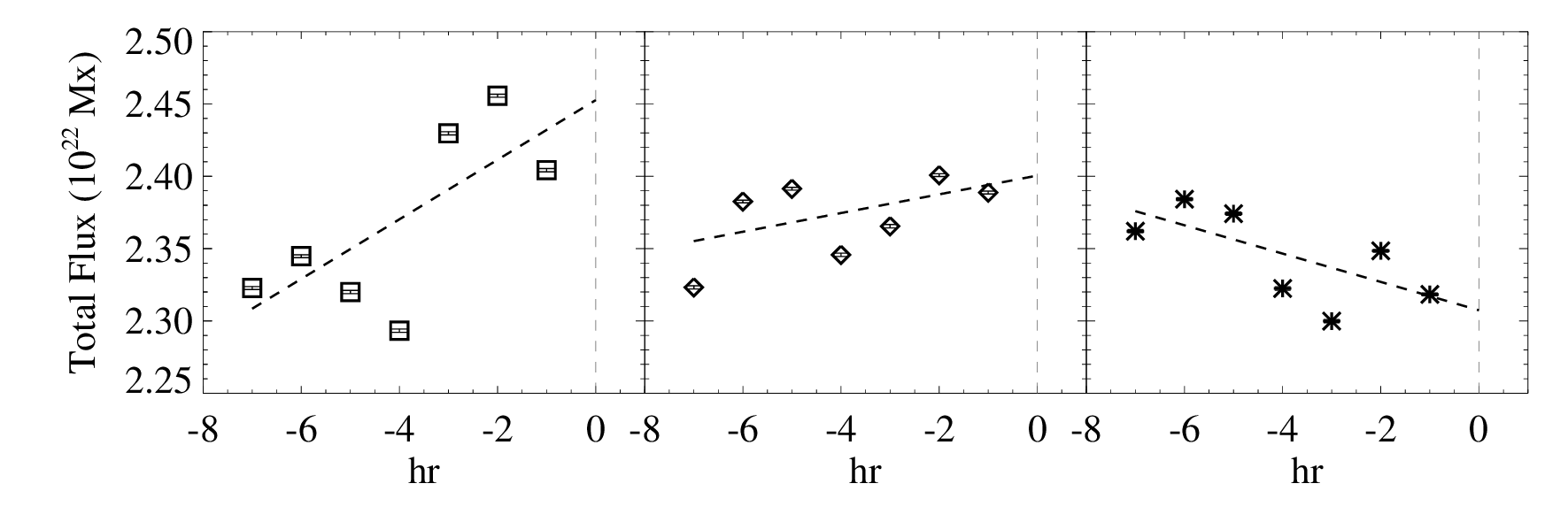} \\
\includegraphics[width=0.6\textwidth,clip, trim = 5mm 5mm 0mm 0mm, angle=0]{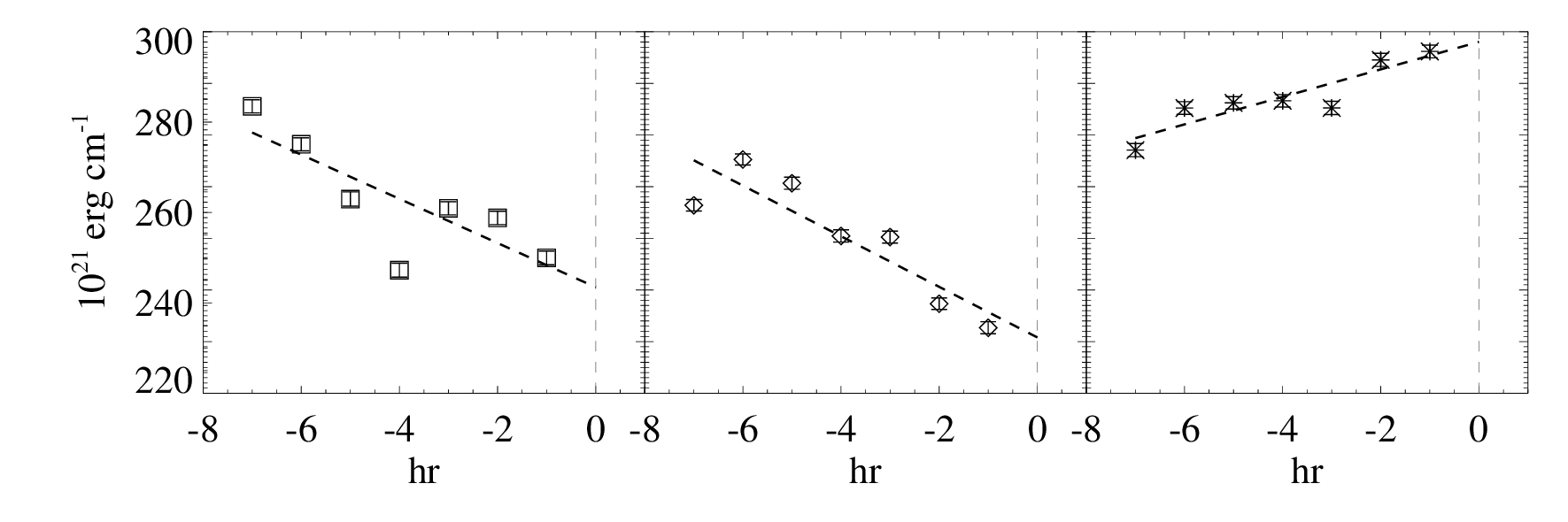} \\
\includegraphics[width=0.6\textwidth,clip, trim = 5mm 5mm 0mm 0mm, angle=0]{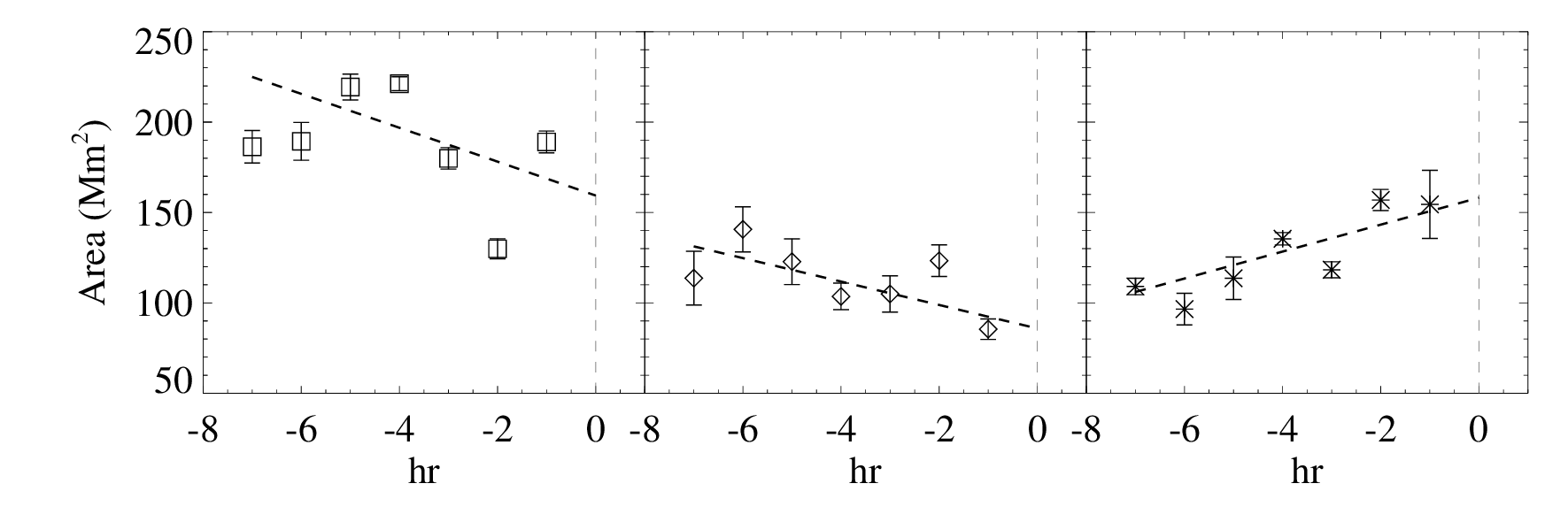} \\
\includegraphics[width=0.6\textwidth,clip, trim = 5mm 5mm 0mm 0mm, angle=0]{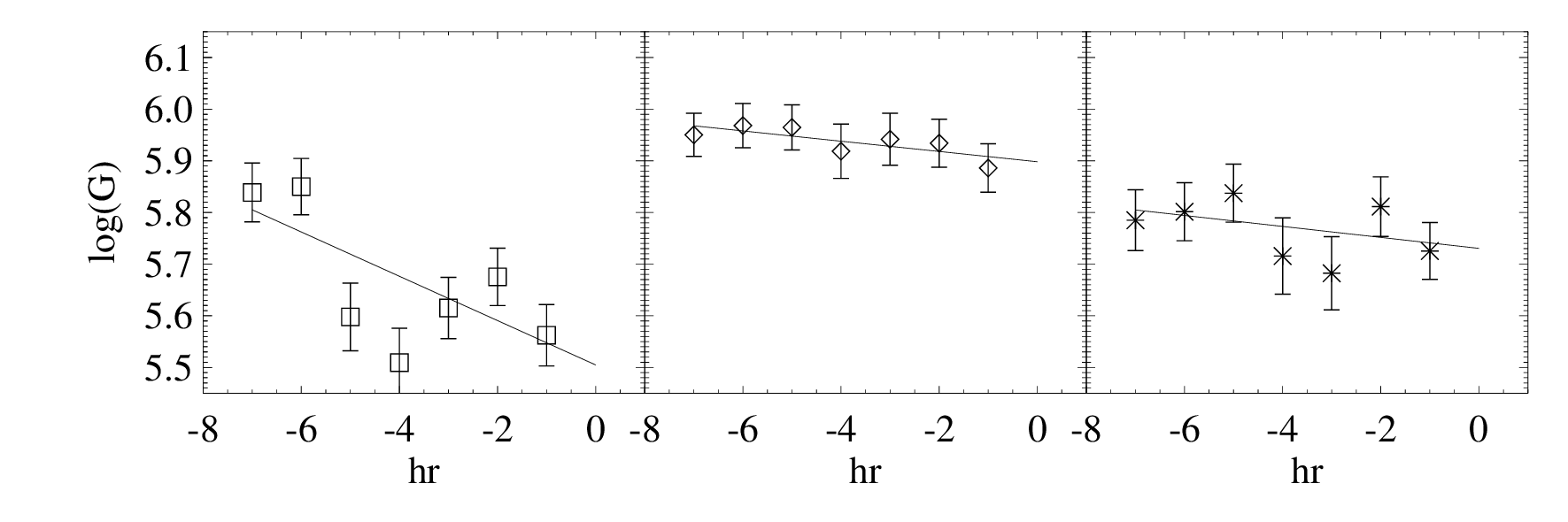}
\end{center}
\caption{Plots of 6-hr intervals (7 samples) of time-series parameters from the 
magnetic field analysis of NOAA~AR\,11283 for three time intervals.
{\bf From Top:} the evolution of the total magnetic flux, 
the evolution of the proxy of the free magnetic energy; 
evolution of the area of strong magnetic shear in the
vicinity of the magnetic neutral line; {\bf Bottom:} the evoluton of the log of the
$\mathcal{R}$ parameter.  The times relative to the issuance times are shown, 
in these cases the three intervals end just before the X1.8 flare on 2011.09.07, just
before the X2.1 flare on 2011.09.06, and a flare-quiet period of 2011.09.7 respectively.
That is, this would be an example invoking a super-posed epoch analysis whereas in a 
forecasting approach the forecast issuance time would likely be a particular time of day.}
\label{fig:preflareplots}
\end{figure}

The \nci\ as implemented for research on flare-imminent active regions is designed 
to include the recent evolution of the
parametrizations discussed in the prior sections.  For the photospheric
magnetic field parameters and the coronal topology parameters,
a linear function is fit to the parameters computed at each of the
seven times acquired.  The slope and intercept at the forecast issuance time are
used (Figure~\ref{fig:preflareplots}) as two separate parameters.  The intercept
is used instead of the mean (or similar) of the times considered, in
order to account for latency between data acquisition, analysis time,
and the forecast issuance time.

The HMI magnetic field data incur a temporal variation as a function of the
orbital velocity of the SDO spacecraft, as described in \citet{hmi_pipe}.
Although efforts are underway to mitigate the impact \citep{Schuck_etal_2016},
for the moment, any analysis based on temporal variation of the magnetic field
must accommodate these variations.  In the case of \daffs\ the ``$d{\rm t}$''
parameters are all calculated using the same part of the orbit, and as such will
present a bias in the magnitude of parameters calculated for all samples,
but will not preferentially select one population over another.  Forecasts or 
classifications which target a different time and use data from a different
time of day (undergoing a different part of the daily orbit) are considered separately.

% entries 66:72 are last 7-hr like we've been doing, on 2011.09.07 (ends w/ X1.8).
% entries 42:48 are 15:48 -- 21:48 on 2011.09.06, just before its X2.XX flare.
% entries 54:60 are 03:48 -- 09:48, quiet and after flare of previous day

\subsection{Discriminant Analysis}
\label{sec:nci_da}

All forecasting methods include a statistical analysis or machine learning in order to
produce a forecast from the observational input.  Ranging from the
simple (a correlation, \citep[{\it{e.g.}}][]{Falconer_etal_2011}) to
quite complex (a Cascade Correlation Neural Network \citep[{\it{e.g.}}][]{Ahmed_etal_2013}),
the basic goal is the same: use a description of past event behavior in
the context of past data, to predict future events given new data.
DA as a general statistical characterization 
was first applied to solar flare forecasting
in an early attempt to quantify improvements which could be made through multi-parameter
analysis \citep{flareprediction}.  

\begin{figure}
\centerline{
\includegraphics[width=0.5\textwidth,clip, trim = 0mm 0mm 0mm 0mm, angle=0]{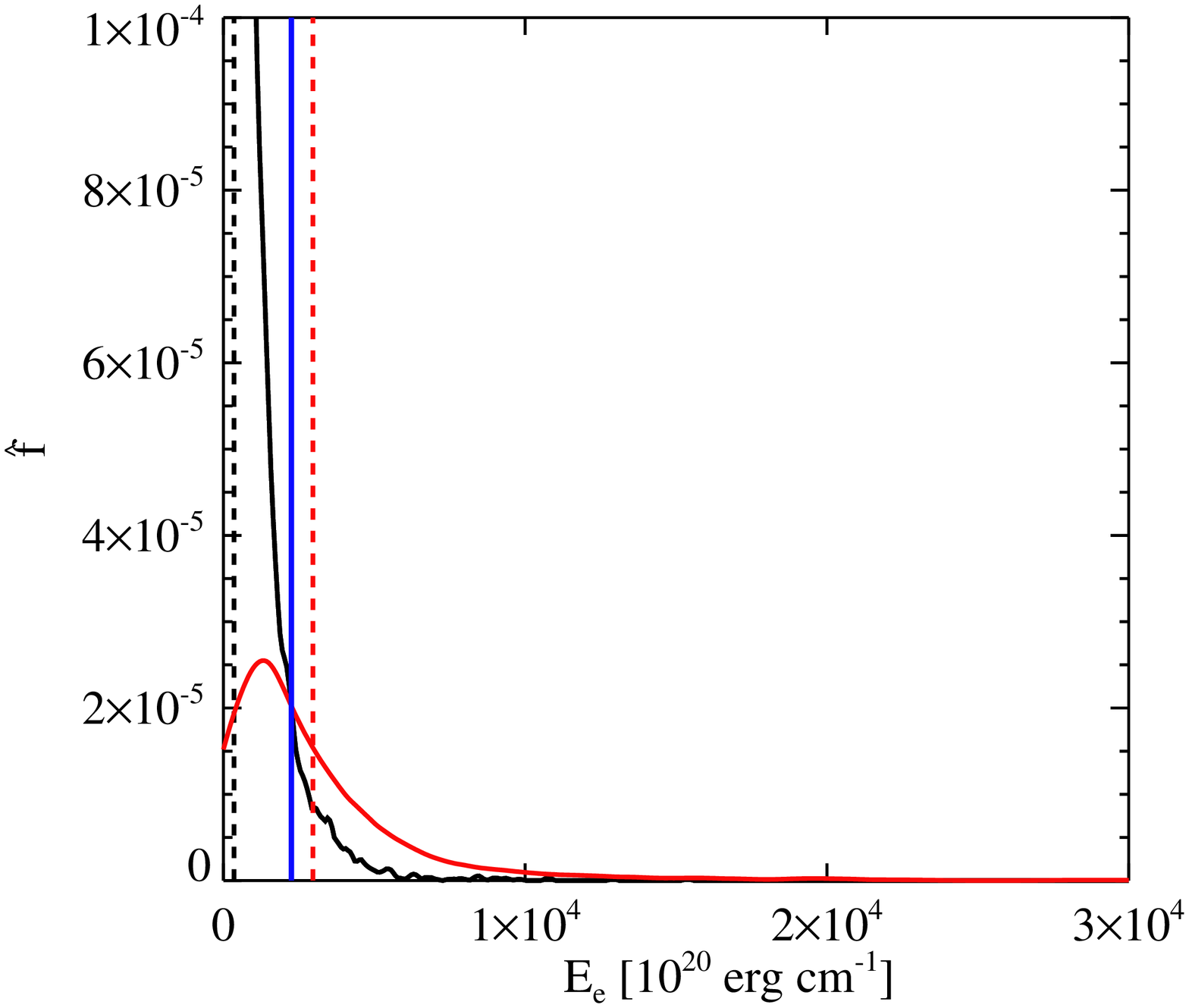}
\includegraphics[width=0.5\textwidth,clip, trim = 0mm 0mm 0mm 0mm, angle=0]{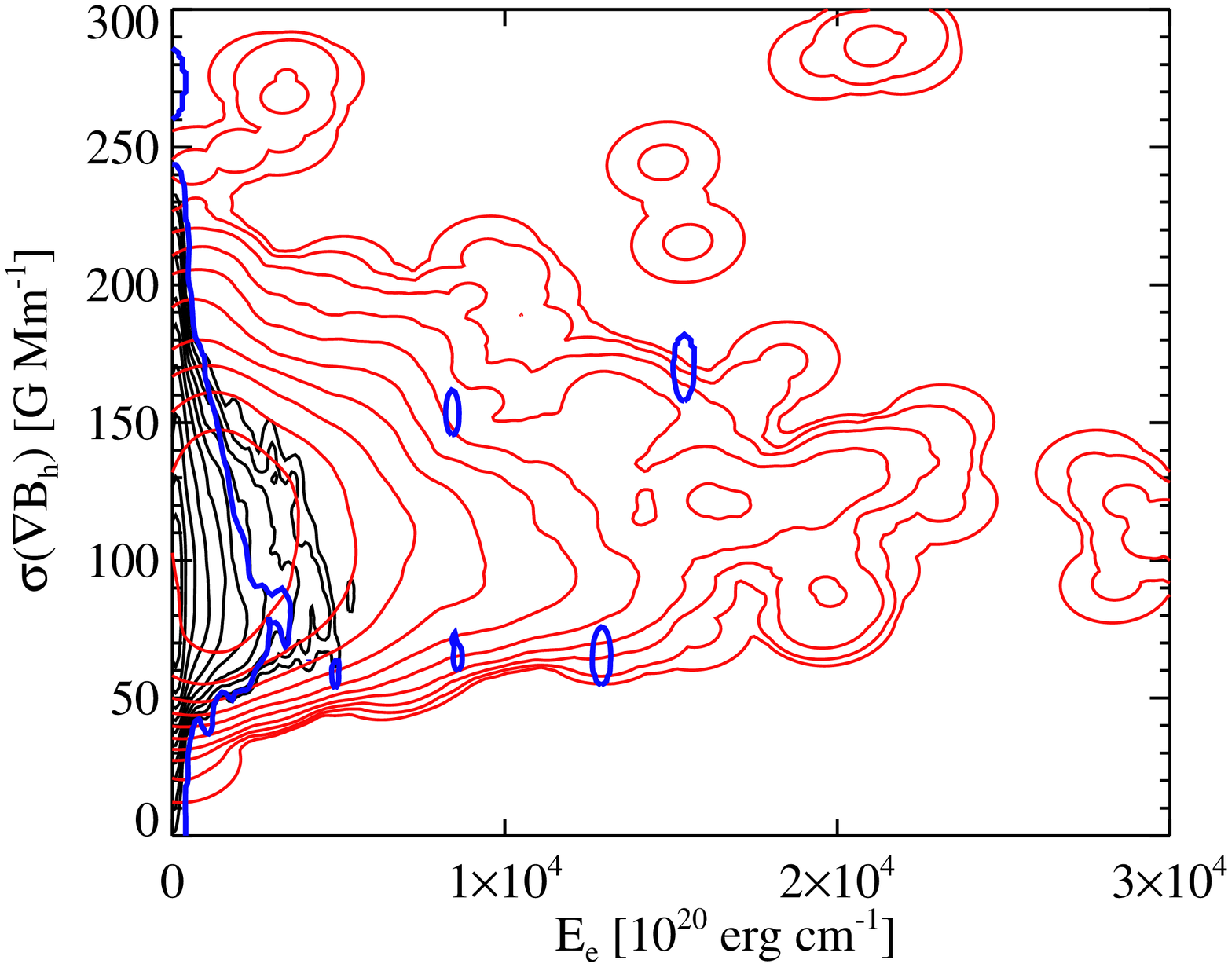}}
\caption{Examples of nonparametric discriminant analysis for one variable (total ``excess'' 
photospheric energy, see \citep{params}) and two variables (same, plus the standard deviation
of the horizontal gradients of the horizontal component of the magnetic field, {\it ibid.})
for the \CC\ event definition.  For both, event / non-event NPDA estimates 
are shown in red/black respectively, and the 50\% probability threshold is 
shown in blue.}
\label{fig:da_plots}
\end{figure}

The \nci\ research into flare productivity generally uses NPDA (\S~\ref{sec:da}).
Examples of 1-variable and 2-variable results are shown
in Figure~\ref{fig:da_plots}, for one event definition (\S~\ref{sec:nci_event_defs}).
The issues raised regarding small tail sample sizes are readily apparent 
in the graphic for the 2-variable sample results.

\subsubsection{\nci\ Flare Research Error Estimation}
\label{sec:nci_error}

The bootstrap method described above (\S~\ref{sec:errors})
is performed randomly on each HARP for region forecasts; uncertainties for
the full disk probabilities are calculated by performing the bootstrap
on daily ensembles.  The resulting skill scores are computed for each
draw, and their uncertainties are calculated by the standard deviation
of the draws.

\subsubsection{Accounting for Statistical Flukes}
\label{sec:nci_flukes}

As mentioned above, a Monte-Carlo approach can be used to estimate
how likely any of the reported scores would be by statistical fluke,
given the sample sizes and number of parameters considered.
Using the flare-probability 
parameters described above and the HMI-data sample sizes,
we estimate there would be $<\!\!1\%$ chance of a resulting
${\rm BSS}\!>\!0.001/0.002/0.003$ by chance alone for
single variable NPDA for {\tt C1.0+/M1.0+/X1.0+} flares, respectively.  

\subsection{\nci\ Flare Research: Evaluation}
\label{sec:nci_eval}

We present first some representative results using the ``research-based''
modules for flare-imminent classification and \nci; this includes
the magnetic parameters, the topological parameters, the prior
flare parameters, and the temporal behavior of each as appropriate
(Table~\ref{tbl:nci_region_metrics}).  We focus on region forecasts,
as that is most appropriate for research purposes.  This report serves
at some level as an update to earlier publications \citep{dfa3,SWJ}, now
with updated data and a significantly larger sample size: data covering
the full {\it SDO} mission for ``definitive'' vector magnetic field HARPs
%are used 2010 May 01 -- 2017 June 30 (as available at the time of this writing).
are used 2010 May 01 -- 2017 June 30 (as available at the time of this writing\footnote{
The NCI demonstration results shown here
include parameters from HMI ``mode-L'' data starting from 2016.04 that were discovered to include 
alignment errors.  As of 
this manuscript's acceptence, the affected data are being reprocessed by the HMI team but
the task has not yet completed.  Substantive quantitative differences are present between 
parameters generated from the initially released model-L data and examples of
reprocessed data.  Especially impacted are parameters that rely on the horizontal 
component of the field.  What is presented here are the results given the data available.}).
This provides almost 30,000 HARP-days (individual HARPs acquired on
separate single days).

\begin{table}
\caption{Representative \nci\ Research-Mode Region-by-Region Flare Classification Performance Metrics}
\begin{center}
\begin{tabular}{lllllll} \hline 
\multicolumn{7}{c}{Region-by-Region, 2010.05.01 -- 2017.05.31 }   \\ \hline
Event &  Event &  Parameter(s) & \RC & \Brier\ & \ApSS\ &  Optimal  \\ 
Def. &  Rate &   & &  &  &  \TSS \\ \hline
{\tt C+1} & 0.0879 & $\varphi_{tot}$, $\sigma(\gradh B_z)$ & 0.937 $\pm$ 0.005 & 0.40 $\pm$ 0.01 & 0.28 $\pm$ 0.01 & 0.68 $\pm$ 0.01 \\
{\tt M+1} & 0.0145 & $E_e$, $\sigma(|h_c|)$ & 0.987 $\pm$ 0.001 & 0.26 $\pm$ 0.02 & 0.14 $\pm$ 0.03 & 0.76 $\pm$ 0.02 \\
{\tt X+1} & 0.0013 & $I_{tot}^h$, $\mathcal{F}(\psi_{NL}>45\degr)$ & 0.9988 $\pm$ 0.0003 & 0.12 $\pm$ 0.06 & 0.12 $\pm$ 0.07 & 0.74 $\pm$ 0.07 \\ \hline
{\tt C+2} & 0.0837 & $\varphi_{tot}$, $\sigma(\gradh B_z)$ & 0.936 $\pm$ 0.007 & 0.35 $\pm$ 0.01 & 0.23 $\pm$ 0.01 & 0.64 $\pm$ 0.01 \\
{\tt M+2} & 0.0134 & $E_e$, $\sigma(|h_c|)$ & 0.990 $\pm$ 0.001  & 0.22 $\pm$ 0.02 & 0.11 $\pm$ 0.03 & 0.69 $\pm$ 0.02 \\
{\tt X+2} & 0.0013 & ${\rm FL}_{\rm 24}$, $\sigma(\Psi_{\rm NL,W})$  & 0.9988 $\pm$ 0.0002 & 0.13 $\pm$ 0.06 & 0.12 $\pm$ 0.06 & 0.69 $\pm$ 0.07 \\ \hline
{\tt C+3} & 0.0767 & $\varphi_{tot}$, $\log(\mathcal{R_{\rm nwra}})$  & 0.938 $\pm$ 0.007 & 0.31 $\pm$ 0.01 & 0.19 $\pm$ 0.01 & 0.60 $\pm$ 0.01 \\
{\tt M+3} & 0.0124 & $I_{tot}^h$, $\overline{\rho_e}$ & 0.9884 $\pm$ 0.0008 & 0.15 $\pm$ 0.02 & 0.07 $\pm$ 0.02 & 0.66 $\pm$ 0.02 \\
{\tt X+3} & 0.0011 & $\log(\mathcal{R_{\rm nwra}})$, $\overline{(r_{ij},\psi)}$ & 0.9989 $\pm$ 0.0002 & 0.11 $\pm$ 0.06 & 0.08 $\pm$ 0.06 & 0.63 $\pm$ 0.10  \\ \hline 
\end{tabular}
\label{tbl:nci_region_metrics}
\end{center}
\vspace{-0.5cm}
{\small Variable symbols can be found in \citet{dfa,dfa2} except 
$\mathcal{F}(\psi_{NL}>45\degr)$ which indicates the fraction of magnetic neutral line
with magnetic shear greater than $45^{\degr}$ and ${\rm FL}_{\rm 24}$ which indicates
the prior flare flux (\S\ref{sec:pff}) for a prior 24\,hr interval.}
\end{table}

For Table~\ref{tbl:nci_region_metrics}, a selection of 
performance results are shown, specifying the parameter combination
used, and some relevant metrics.  \nci\ by default 
uses $P_{\rm th}=0.5$, and the ``well-performing'' combinations are generally 
selected by high \Brier\, as probabilistic forecasts are the most widespread 
in operational settings and a preferred metric for NOAA/SWPC evaluation.
This threshold is used for the quoted \RC\ and \ApSS.
The ``Optimal \TSS'' is the \TSS\ for
which $P_{\rm th} = {\rm Event Rate}$, which is not necessarily the
highest \TSS\ but is generally very close, especially once error bars are 
considered \citep{Bloomfield_etal_2012,allclear}.
The full contingency tables are not presented here, since their entries are sensitive
to $P_{\rm th}$, but all information needed to construct them for 
any chosen $P_{\rm th}$ is provided\footnote{Please see Supplementary Material.}

In Figure~\ref{fig:nci_roc} the Relative (or Receiver) Operating Characteristic Curves (`ROC')
are shown for the entries in Table~\ref{tbl:nci_region_metrics}.  As described in 
Section~\ref{sec:metrics}, a perfect Gini coefficient or ROC Skill Score results in $G1=1.0$, 
manifest by an ROC curve consisting of three points: [0,0], [0,1], and [1,1].
The discontinuities are caused at small probability levels due to many regions being assigned 
the same probability (specifically, the value of climatology).
We also indicate the maximum \TSS\ which is found by stepping through $P_{\rm th}$ values,
with cross-validation but without bootstrap (leading to some expected 
discrepancies with Table~\ref{tbl:nci_region_metrics}), and the Gini coefficient.
Of note is a degradation, but not a substantial one, between increasing latencies.

\begin{figure}
\centerline{
\includegraphics[width=0.33\textwidth,clip, trim = 0mm 0mm 0mm 0mm, angle=0]{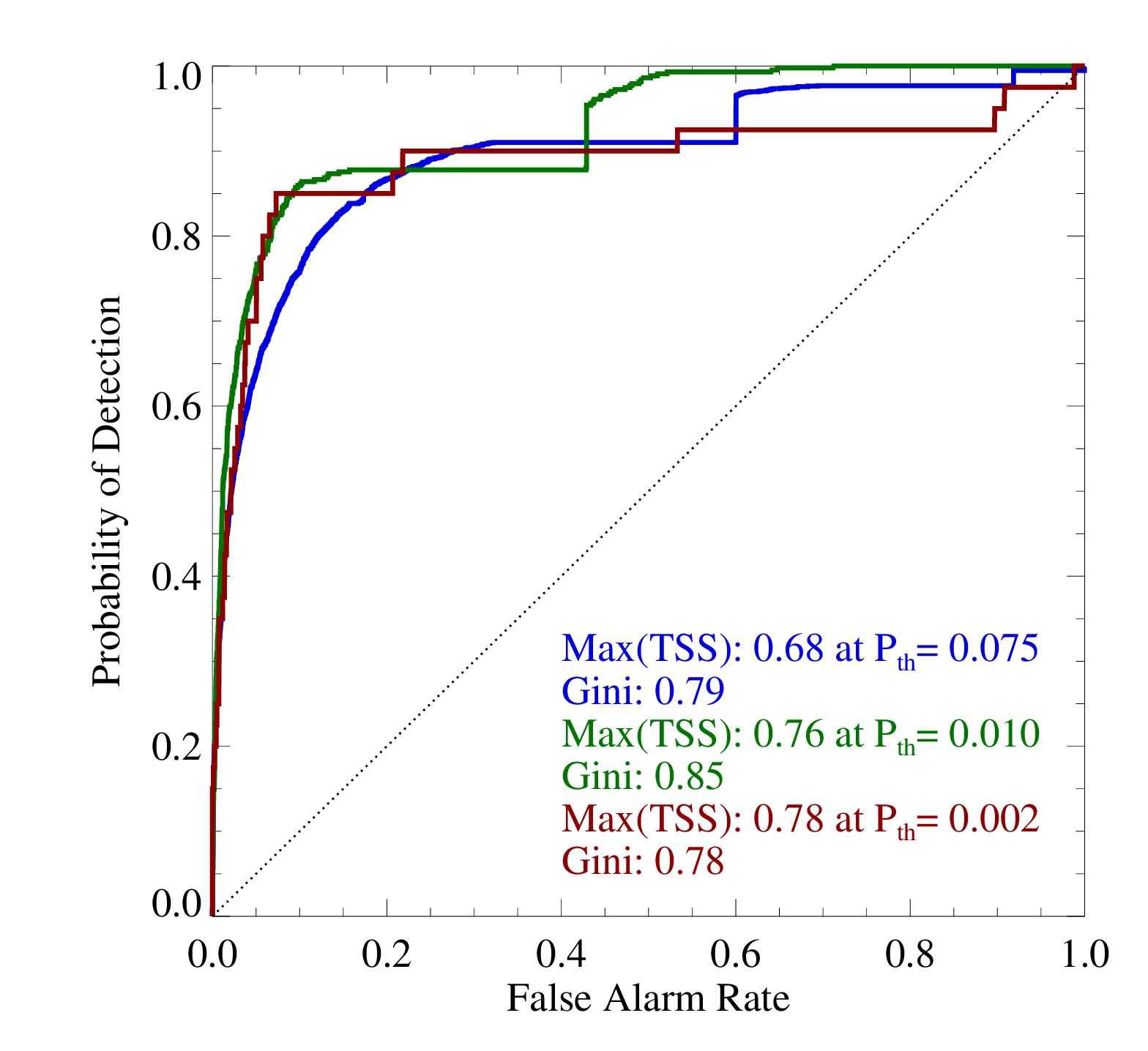}
\includegraphics[width=0.33\textwidth,clip, trim = 0mm 0mm 0mm 0mm, angle=0]{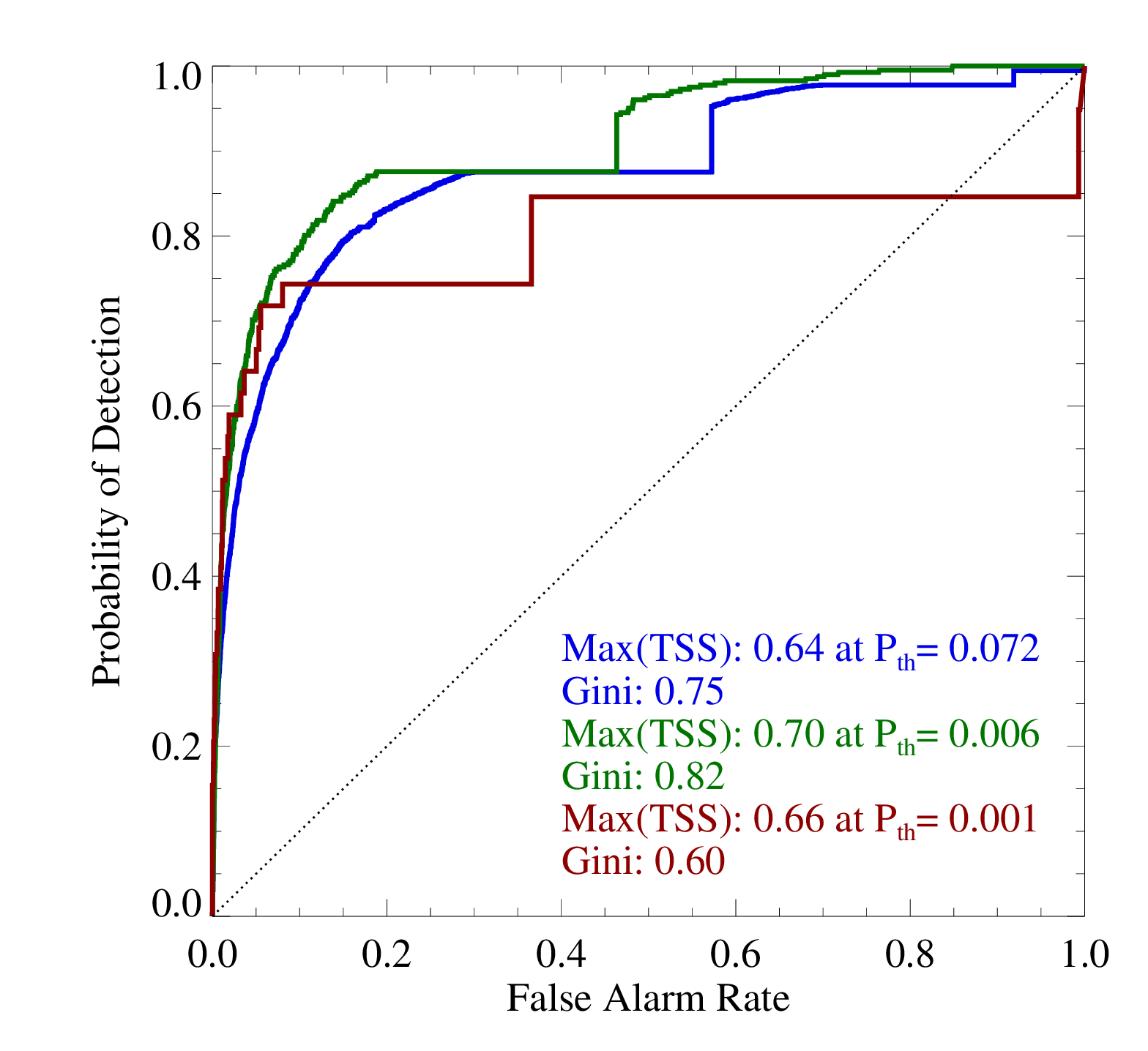}
\includegraphics[width=0.33\textwidth,clip, trim = 0mm 0mm 0mm 0mm, angle=0]{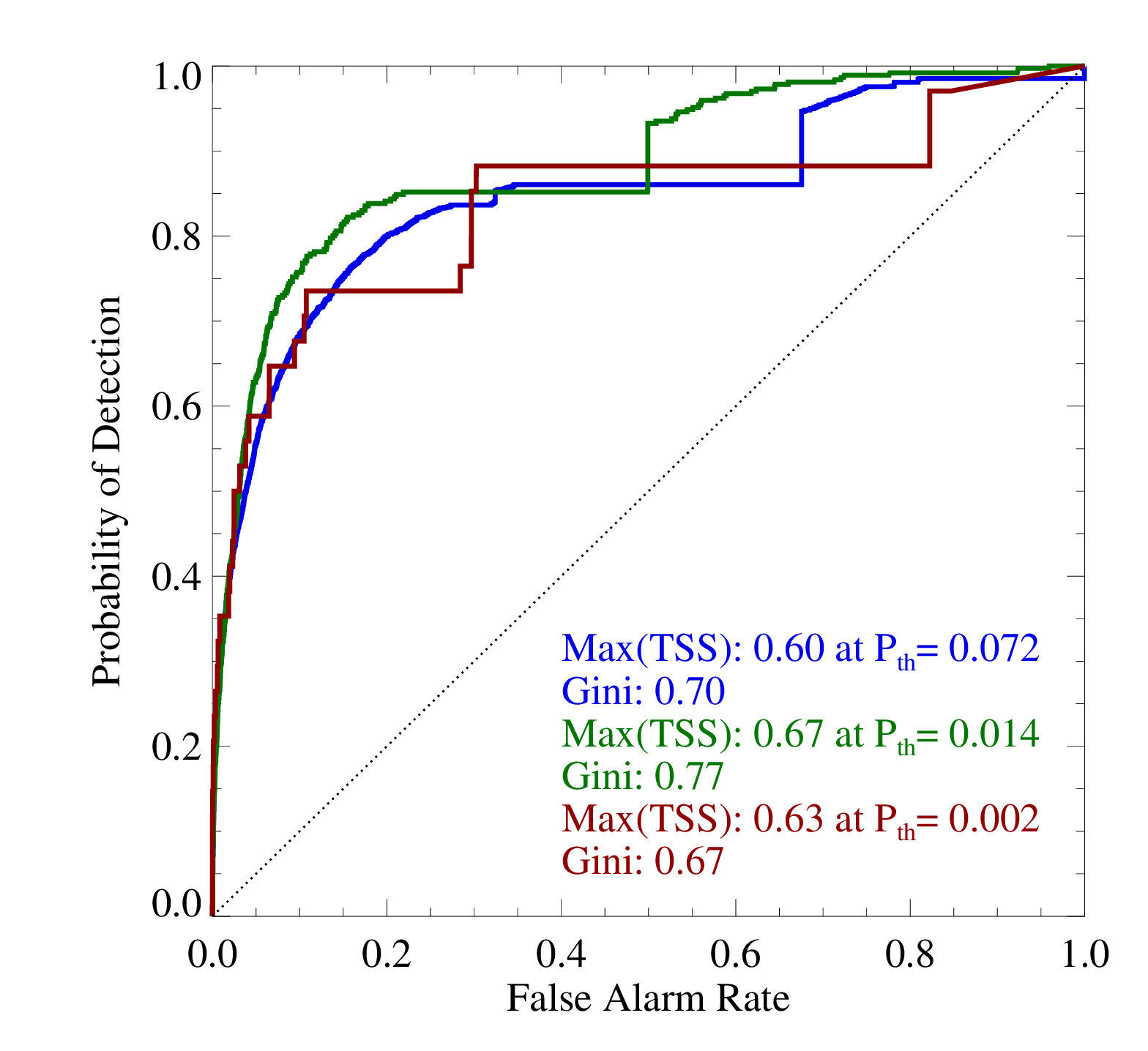}}
\caption{Relative (or Receiver) Operating Characteristic Curve for \nci\ research on 
flaring {\it vs.} flare-quiet active regions, using a forecasting context as described in the text.
Blue/Green/Red curves indicate the three event definitions (\CC, \MM, \XX) for effectively 
0\,hr, 24\,hr, and 48\,hr latencies (left:right), region-by-region ``forecasts'' using the 
parameter combinations for each event definition listed in Table~\ref{tbl:nci_region_metrics}.
Also noted on each plot are the maximum \TSS\ (TSS) achieved, the probability 
threshold $P_{\rm th}$ used for that maximum TSS score, and the Gini coefficient for the relevant curve.}
\label{fig:nci_roc}
\end{figure}

\subsection{Flare Research: Results}
\label{sec:nci_results}

In general, we find that the top-performing parameter pairs routinely include (but are not 
exclusive to) the following parameters categories (in no particular order):
\begin{itemize}
\item A measure of recent flare activity ({\it e.g.} ${\rm FL}_{\rm 12}$ or ${\rm FL}_{\rm 24}$),
\item A measure of size ({\it e.g.} $\Phi_{tot}$, $I_{tot}$),
\item A measure of energy storage and non-potential magnetic field ({\it e.g.} 
$\log(\mathcal{R}_{\rm nwra})$, $\mathcal{F}(\Psi_{NL} > 45^\circ)$),
\item A measure of magnetic complexity ({\it e.g.} $\overline{\varphi_{ij}}$, $\kappa(B_z)$)
\end{itemize}

\noindent
Often a parameter based on the temporal evolution is included, however
within the error bars there are always combinations without temporal
evolution which perform similarly.  Using two-parameter NPDA, we generally
find dozens of parameter combinations that perform similarly within the
error bars.  When looking at some of the better-performing combinations,
we find a gradual decrease in performance for probabilistic forecasts
with respect to latency, and a substantial decrease in performance with
increasing event magnitude.

These results are consistent with both our earlier results \citep{dfa,mct,dfa3,SWJ} and with
the present state of the literature \citep[{\it{e.g.}},
][]{Falconer_etal_2014,BobraCouvidat2015,Al-Ghraibah_etal_2015,Nishizuka_etal_2017,Murray_etal_2017},
but should not be compared directly due to different event definitions,
testing intervals, and validation methodologies \citep[{\it c.f.}
][]{allclear}.

The {\it lack} of a few well-identified parameters which definitively distinguish the 
two defined populations highlights the challenge of using statistical empirical 
relationships to investigate the fundamental physics of flares.  Flares occur
in regions which are large and magnetically complex, and preferentially in 
regions which have flared previously.  The latter point is consistent with 
flare models based on Self Organized Criticality \citep{Lu_Hamilton_1991,Strugarek_etal_2014}.  
However, one can also view some of the empirical results as guidance
for modeling efforts, which (thus far) rarely require that the boundary
field to provide distinguishing differences of magnetic complexity in addition
to a sheared polarity inversion line \citep[although see ][]{Kusano_etal_2012}.  

\section{The \daffs\ Near-Real-Time (NRT) Flare Forecasting Tool}
\label{sec:daffs}

\begin{figure}
\centerline{\includegraphics[height=0.98\textwidth,clip, trim = 20mm 5mm 45mm 10mm, angle=-90]{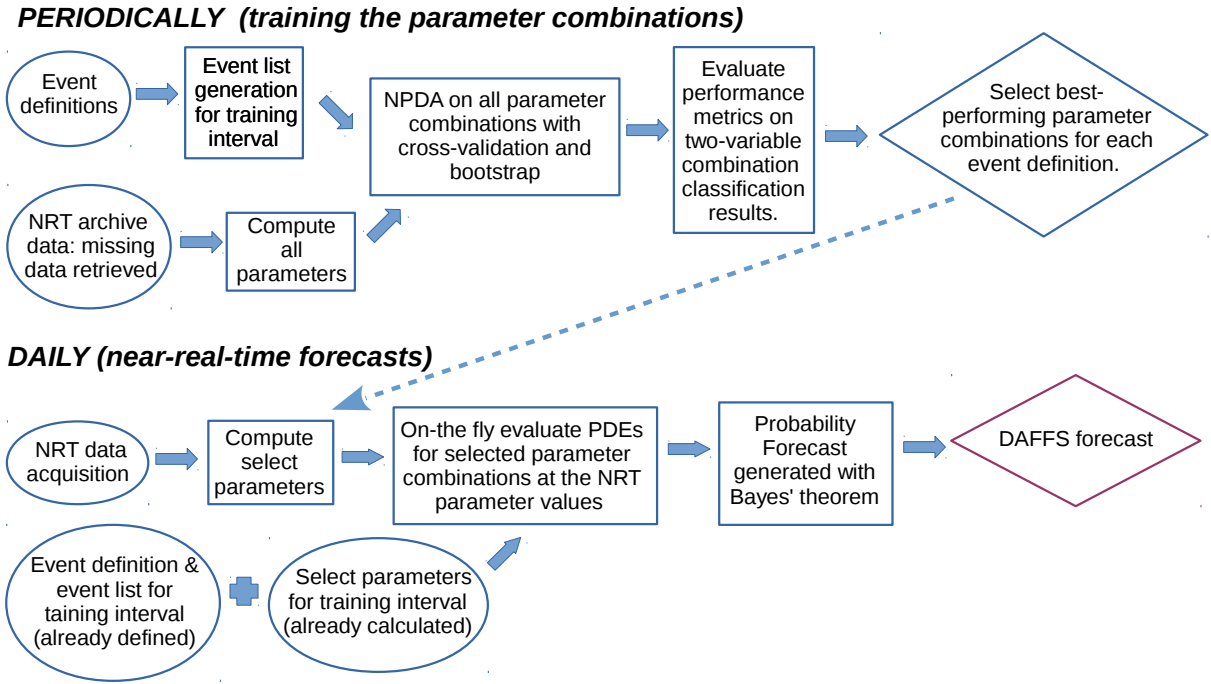}}
\caption{A very generalized flow chart for the \daffs\ system, outlining training to 
determine best-performing parameter pairs at the top, whose results impact the daily near-real-time
operation, in the bottom flow chart.  Circles generally 
indicate input, squares are processes, and diamonds are output. }
\label{fig:flow_daffs}
\end{figure}

The \nci\ infrastructure recently bifurcated to include
a near-real-time operational flare forecasting tool.
The Discriminant Analysis Flare Forecasting System (\daffs)
is the result of a NOAA Small Business Innovative Research (SBIR) Phase-II
contract to NWRA.  To achieve a truly operational forecasting tool, many aspects that
originated from the \nci\ were redesigned for automated
stand-alone performance (no ``human in the loop''), with operational redundancy.  In short,
the \daffs\ {\tt cron} scripts use the prescribed
forecast issuance time as the basis for their schedules, with drivers of all needed modules written in Python.
There are a few key differences from the research-based \nci\ approach above,
including some design features not yet implemented due to limited resources,
and we describe those below.  A generalized flow-chart for \daffs\ is provided
in Figure~\ref{fig:flow_daffs}.

\subsection{NRT Data Sources}
\label{sec:nrt_data}

\begin{figure}
\centerline{\includegraphics[width=0.5\textwidth,clip, trim = 10mm 0mm 0mm 5mm, angle=0]{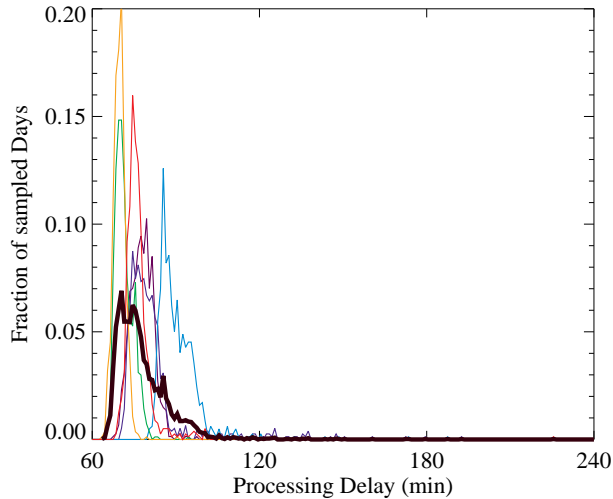}}
\caption{Histogram of processing delay distributions (elapsed time in between data 
acquisition and its availability for transfer) for six fairly random weeks between 2014 and 
2015 (colored lines) and the distribution for all points (black).  While there is a tail 
to delays greater than 90 minutes, the majority of data are available within that time.}
\label{fig:nrt_timing}
\end{figure}

The first difference is the source of the vector magnetic field data.
For the research system the HMI ``definitive'' series are used, but
for the operational forecasts, the HMI ``Near Real Time'' (NRT) data are
used.  Specifically the {\tt hmi.ME\_720s\_fd10\_nrt} full-disk
vector field data are retrieved, along with the NRT HARP information
from the {\tt hmi.Mharp\_720s\_nrt} series and the disambiguation
results from the {\tt hmi.Bharp\_720s\_nrt} series.  The expected
latency for the HMI NRT data processing and retrieval was investigated
(Figure~\ref{fig:nrt_timing}), and the estimates for expected delays
are such that the latest target data practicable are slightly more
than 2hr prior to forecast issuance time;  this imposes a latency by
default, as summarized in Table~\ref{table:nrt_data_source_timing}.
If the target data are unavailable due to processing or transfer delays,
then the target is moved back and forth in time by the HMI vector field cadence
(12 min) up to one hour prior and including up to 48 min later -- beyond
which it is deemed to be missing data (see Table~\ref{tbl:nrt_timing}).
The timeline for producing a near-real-time \daffs\ forecast is summarized
in Table~\ref{tbl:nrt_timing}.

\begin{table}
\caption{\daffs\ Near-Real-Time Data Target Timing}
\centering
\begin{tabular}{cll} \hline
\multicolumn{3}{l}{As relevant for a ``midnight'' forecast:}   \\
Time & Data Source & Keyword \\ \hline
21:48:00 & HMI/MAG target time  = Master - 2.1h & {\tt DAFFS\_HMI\_LATENCY} \\
22:30:00 & NOAA GOES/PFF target time = Master - 1.4h & {\tt DAFFS\_PFF\_LATENCY} \\
22:54:00 & GONG/LOS\_MAG target time = Master - 1.0h & {\tt DAFFS\_GONG\_LATENCY} \\
23:54:00 & Master time & \\
00:00:00 & Forecast time = Master + 0.1h & {\tt DAFFS\_TFORECAST\_LATENCY} \\ \hline
\end{tabular}
\label{table:nrt_data_source_timing}
\end{table}

\begin{table}
\caption{\daffs\ Data Acquisition and Processing Timeline}
\centering
\begin{tabular}{cl} \hline
\multicolumn{2}{l}{As relevant for a ``midnight'' forecast:}   \\
Time & Task \\ \hline
23:34 & Retrieve GONG NRT data, targeting 22:54. \\
23:36 & Retrieve data from HMI magnetogram series ({\tt hmi.M\_720s\_nrt}) for full-disk context\\
23:36 & Retrieve NRT HMI full-disk data ({\tt hmi.ME\_720s\_fd10\_nrt}) via NetDRMS, and \\
 & extract patches using {\tt hmi.MEharp\_720s\_nrt, hmi.Hharp\_720s\_nrt} series.\\
&  Attempt 21:48:00 target record; if it does not exist, wait one minute \& retry.   \\
(21:37) & If target record still does not exist, query for data in the following \\
& order retrieve closest available record within [-60,+48] of target: \\
& 21:36, 22:00, 21:24, 22:12, 21:12, 22:24, 21:00, 22:36, 20:48 \\
23:37 & Query E-SWDS for latest SWPC NRT flare events, AR assignments; simultaneous  \\
 & {\tt ftp} query \& transfer \\
23:54 & 0. Plot HARPs on full-disk $\Bl$ image, for \daffs\ landing page (Figure~\ref{fig:nrt_daffs})\\
      & 1. Update NWRA database with latest events via NOAA E-SWDS database query\\
      & 2. Update NRT HARP/NOAA translation table in NWRA database \\
      & 3. Extract GONG patches via NOAA E-SWDS database query of visible active regions \\
      & 4. Generate parameters, forecasts \\
      & 5. Link the main webpage to the new forecast \\ \hline
\end{tabular}
\label{tbl:nrt_timing}
\end{table}

The needed NOAA data (up to date flare events and coordinates of visible
numbered active regions) are retrieved through queries to {\tt ``E-SWDS''}.
The needed data are also retrieved from the public {\tt ftp} service
as backup.

The GONG data are retrieved from the NSO near-real-time dissemination
page, {\url{https://gong2.nso.edu/oQR/zqa/}}; searches are performed via Python script 
to find all sources (magnetogram images from all GONG sites) that exist for a given day 
and targeted time.  The choice of target time (Table~\ref{table:nrt_data_source_timing})
is in part due to GONG data being published ``on the fours'', and (similar to HMI NRT data)
the time by which the data would be reliably made available.
%http://gong2.nso.edu/oQR/zqa/201512/bbzqa151208/bbzqa151208t2154.fits.gz
All matching gzipped FITS files are downloaded for evaluation (see \S~\ref{sec:blos}),
and the image with the best seeing is then used.

The target times for data sources were chosen according to when
the target data products typically become available for transfer
(Table~\ref{table:nrt_data_source_timing}).  The master time was
chosen to be 6 minutes (0.1h) prior to forecast time, to give the
forecast code sufficient time to complete by the forecast issuance time.
By referring all of the data acquisition and processing time to a master
time (and specifically a master time which is on the same day as the
data acquisition), and setting the relative times through keywords,
flexibility is afforded for setting different forecast issuance times.

Timing tests were performed for each aspect of the pipeline, from data
retrieval and staging to producing a forecast and making it live, to
come up with a task schedule (implemented via {\tt cron}) for smooth
and automatic operation as well as automated failure handling.  In the
case of HMI, if the target time is not available, adjacent times are
searched as described in Table~\ref{table:nrt_data_source_timing}.
The HMI NRT data are generally processed and available for transfer
within 90 minutes of the observation time, but delays are not unusual
(see Figure~\ref{fig:nrt_timing}).  If no HMI sources can be found after
a certain amount of time, the forecast is presently issued based on
evaluating NOAA-provided flare history parameters (see \S~\ref{sec:pff}).
GONG data are intended to serve as a backup for HMI but some work remains
to fully integrate this parallel system.

\begin{figure}[t]
\centerline{
\includegraphics[width=0.50\textwidth,clip, trim = 0mm 0mm 0mm 0mm, angle=0]{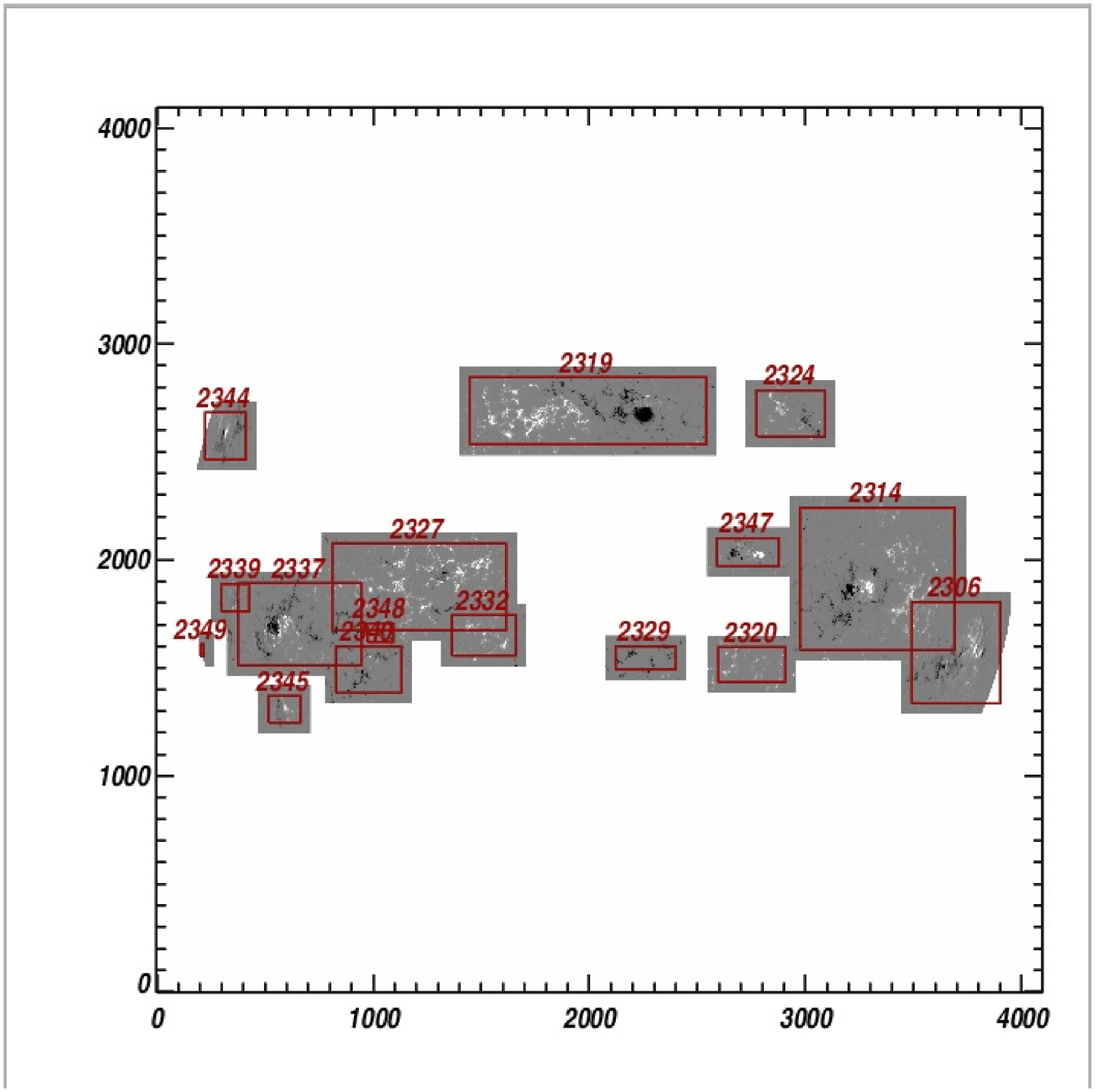}
\includegraphics[width=0.50\textwidth,clip, trim = 0mm 0mm 0mm 0mm, angle=0]{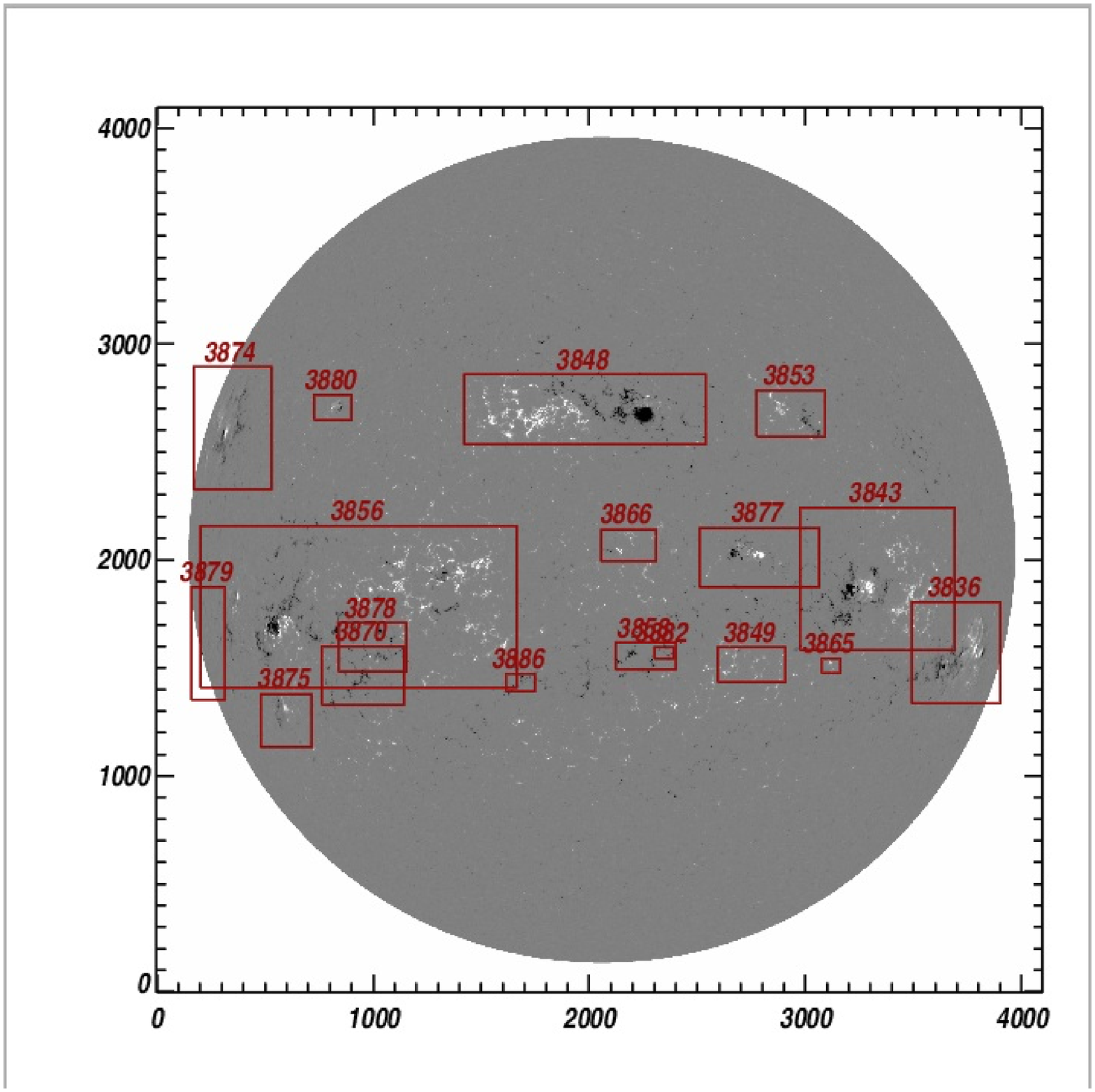}}
\caption{Full-disk line-of-sight magnetograms from SDO/HMI for 2014.03.18 21:48 TAI.
{\bf Left:} Near-Real-Time data showing the near-real-time HARPs and their numbering,
{\bf Right:} Definitive data, showing definitive HARP boxes and their numbering.
Of note are the differences in coverage and in regions in the near-real-time 
which are merged for definitive HARPs.  For both NRT and definitive HARPs, boxes
may overlap but the ``active pixels'' ({\it c.f.} Fig.~\ref{fig:11283gram}) will not.}
\label{fig:nrt_definitive_data_diffs}
\end{figure}

Differences between the SDO/HMI definitive and near-real-time data
arise at a few steps in the data reduction, and some are demonstrated
in Figure~\ref{fig:nrt_definitive_data_diffs}.  Of note, not all of the
full-disk is inverted for the NRT release (only the NRT-HARP areas
plus a buffer); the disambiguation is performed with a faster cooling
schedule.  For any specific area, there may be detectable differences
on a pixel-by-pixel basis, however there is no systematic under- or
over-reporting of magnetic field strengths, azimuthal angle differences,
or noise levels \citep{hmi_sharps}.  Statistically, the distributions
of resulting parameters generally agree well between the NRT and definitive
HMI data series.

Additionally, the NRT HARP definitions themselves are generated in near
real time, and do not have the benefit of size or identity consistency
over a disk-passage as is the case for the definitive HARP series 
(Figure~\ref{fig:nrt_definitive_data_diffs}). This
difference actually precludes exact region-by-region comparisons between 
the research \nci\ results and NRT \daffs\ results, although comparisons 
are possible via full-disk forecasts.

\subsection{NRT \daffs\ Implementation Specifics}
\label{sec:nrt_implement}

For both the HMI NRT and the GONG data for \daffs , only a subset of parameters
are considered during training ($<50$\% of the full science-investigation
list) in order to remove redundancy and highly correlated variables (for
example, separately the positive magnetic flux, the negative magnetic
flux, and the total magnetic flux).  At present, the MCT module is not
used for the NRT \daffs, and data are only examined at a single time;
no ${\rm d}X/{\rm d}t$ analysis is included.

Metadata from {\tt hmi.Mharp\_720s\_nrt} is used to
match HARPs to AR numbers.  For later training efforts, back-propagation 
of HARP/NOAA matches is performed for those HARPs that only ever have a single NOAA number 
assigned (to account for the frequent delay of NOAA number assignments). 

\daffs\ runs autonomously twice daily by default, producing
forecasts issued just before 00:00UT and 12:00UT for the event definitions 
and validity periods listed
in Table~\ref{tbl:eventdef}.  Training occurs over a set period
of time, generally using as much NRT data as are available, {\it e.g.}
over the full HMI NRT-data availability period of 2012.10.01 through a
``recent'' month.  The top-performing parameter pairs are then used
for forecasts, separate pairs for each event definition.  
The NPDA estimates of the distributions are re-computed on-the-fly,
against which new data are compared and thus for which flare probabilities
are computed.   

Training occurs separately for permutations of parameters based on those
available: {\it e.g.}, SDO/HMI + NOAA/SXR events, NOAA/SXR events by
themselves, and eventually GONG + NOAA/SXR events.  For the larger events,
single-variable DA and linear DA were considered as well as multi-parameter NPDA,
because of the often improved performance by these simpler approaches
for small sample sizes, however they are not presently used.  The parameters 
presently used (as of this writing) are listed in Table~\ref{tbl:daffs_params};
these are the parameters used for the results reported here.
The exact parameter combinations are determined by the training interval used (indicated), 
and may change upon retraining the system, although as mentioned above there are 
common parameter ``families'' that routinely appear in the top-performing results.

\begin{table}
\caption{\daffs\ Forecasting Parameter Combinations (Training Interval: 2012.10.01 -- 2016.03.31)}
\centering
\begin{tabular}{cl} \hline
Event Definition & Parameter Combination \\ \hline
{\tt C+1} & $\overline{\rho_e}$, $I_{tot}$    \\
{\tt M+1} & $\sigma(\rho_e)$, $I_{tot}^h$   \\
{\tt X+1} & ${\rm FL}_{\rm 12}$, $\varsigma(B_z)$  \\ \hline
{\tt C+2} & $\Phi_{\rm tot}$, $\log(\mathcal{R}_{\rm nwra})$     \\
{\tt M+2} & $\overline{B_z}$, $\log(\mathcal{R}_{\rm nwra})$    \\
{\tt X+2} & ${\rm FL}_{\rm 24}$, $\varsigma(B_z)$     \\ \hline
{\tt C+3} & $\overline{B_z}$, $I_{tot}^h$  \\
{\tt M+3} & $\overline{B_z}$, $I_{tot}$  \\
{\tt X+3} & ${\rm FL}_{\rm 12}$, $\overline{B_z}$     \\ \hline
\end{tabular}
\label{tbl:daffs_params}
\end{table}

%   'C': {"0.0" : ['EFREE_MEAN','JZ_TOT'],
%        "24.0" : ['FLUX_TOT','LIKSR'],
%        "48.0" : ['BZ_MEAN','HC_TOT']},
%   'M': {"0.0" : ['EFREE_STDEV','HC_TOT'],
%        "24.0" : ['BZ_MEAN','LIKSR'],
%        "48.0" : ['BZ_MEAN','JZ_TOT']},
%   'X': {"0.0" : ['FL_12','BZ_SKEW'],
%        "24.0" : ['FL_24','BZ_SKEW'],
%        "48.0" : ['FL_12','BZ_MEAN']}}

\begin{figure}
%\centerline{\includegraphics[width=0.70\textwidth,clip, trim = 0mm 0mm 0mm 0mm, angle=0]{DAFFS_frontspiece_eg3.eps}}
\centerline{\includegraphics[width=0.70\textwidth,clip, trim = 0mm 0mm 0mm 0mm, angle=0]{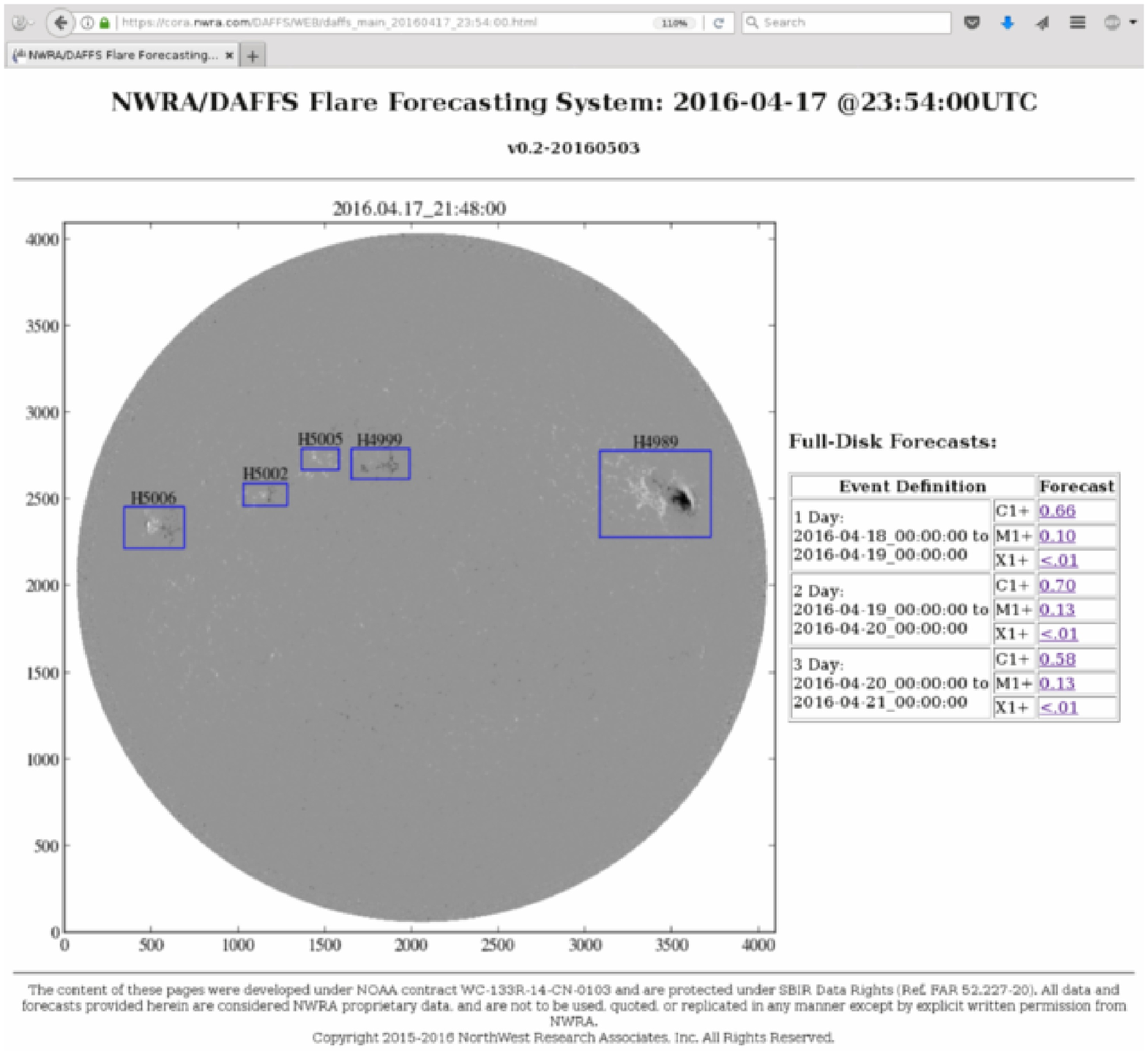}}
%\centerline{\includegraphics[width=0.70\textwidth,clip, trim = 0mm 0mm 0mm 0mm, angle=0]{DAFFS_context_plots_eg3.eps}}
\centerline{\includegraphics[width=0.70\textwidth,clip, trim = 0mm 0mm 0mm 0mm, angle=0]{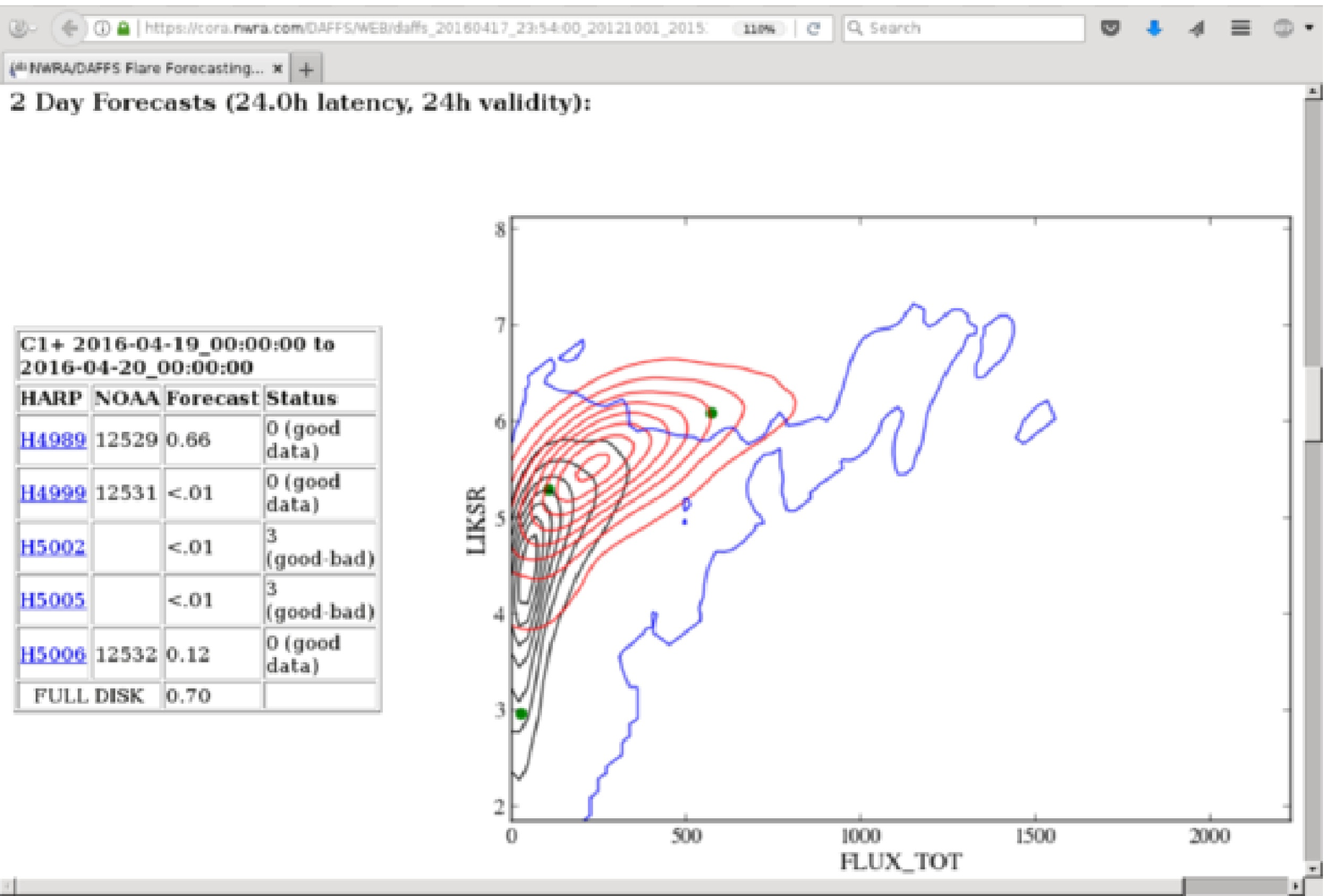}}
\caption{{\bf Top:} ``Landing'' page for the NRT \daffs\ showing a
full-disk line-of-sight magnetogram and defined HARPs for context,
and the full-disk forecasts.    Also shown: the version of the code
running and the training interval being used.  {\tt Bottom:} a context plot
for the {\tt C1.0+}, 24\,hr, 24\,hr latency, region-by-region forecasts, contour indicate the event
(red) / non-event (black) training-sample distributions, the blue contour
is the 50\% forecast level, and the green dots indicate the location of the recent
data on these distribution plots.  The left-hand table shows the 
HARP number, the corresponding NOAA AR(s) as appropriate, the forecast,
and the status flags for the data.  The full \daffs\ forecast for this
date is available for readers through the demonstration page at {\tt www.nwra.com/DAFFS\_home/}.} 
\label{fig:nrt_daffs} 
\end{figure}

Automated graphical output is generated of the locations of the new data
within the parameter space for each event definition (see Figure~\ref{fig:nrt_daffs}).
This allows the user to understand and confirm the context for the given 
probability forecast, and enables a user to track movement of a particular
active region over time (helping to gauge increasing or decreasing flaring
probability).

Performance metrics are generated on demand (see \S~\ref{sec:daffs_results}), as
is re-training the full system (which potentially results in different
parameter pairs being used from thence forward).

\subsection{NRT \daffs\ Redundancy and Operational Details}

An operational system is only as good as its performance when everything
fails.  While there is still significant room for improvement, the
following operational aspects are, or are designed to be, part of the
NRT \daffs\ implementation:

\paragraph{\it No forecast outages:}  A forecast is always issued.  If all 
redundancies fail, climatology is used.

\paragraph{\it Hardware redundancy:}  Not yet implemented.

\paragraph{\it SDO/HMI data outages:} Periodically, there are delays in the SDO/HMI
data processing such that no applicable NRT data (see Table~\ref{tbl:nrt_timing})
are available for a forecast.  In this case, then ``misssing'' data values are assigned and 
a PFF-based forecast with ${\rm FL}_{\rm tot}$ and ${\rm FL}_{\rm 24}$ is prepared.

\paragraph{\it E-SWDS data outage:} The situation of a complete and
long-term E-SWDS data outage likely implies larger problems.  E-SWDS uses
a fail-over server for any outages, and the NWRA database connection
will attempt to connect to it if the primary server is not responding.
The event lists and active-region locations are additionally automatically
updated by a separate {\tt ftp}-based {\tt cron} job from the NOAA
public postings.  Of note, NOAA flare forecasts are also retrieved for
later evaluation comparisons.

\paragraph{\it GONG data outages:} With a world-wide 6-station network,
data outages from GONG are quite rare.  When this does occur, following
the protocol for HMI outages, ``missing'' data values will be assigned
and a PFF-based forecast prepared.

\paragraph{\it Retraining:} the system is retrained on demand, and new
variable-combinations can be employed for the forecasts at that point if
so desired.  This may impact forecasts as results can vary significantly
according to the climatology of the training set, especially for larger
and rarer events.

\paragraph{\it Redundant forecasting:} In order to ensure robustness
of the forecasts, \daffs\ is designed to consider and report on multiple
top performing models; in the case of 2-parameter combinations,
forecasts from the three top performing combinations which contain
unique parameters would be reported.  This would provide essentially
an ensemble forecast that considers 6 parameters, albeit without the
sample-size requirements of training a true 6-variable NPDA forecast.
Although not yet fully implemented, this approach also improves the odds
of successful bad-data rejection.

\paragraph{\it Unassigned Flares} are flare events not assigned to a 
particular active region.  Fairly rare for large events (except when
they occur behind the limb), they are most frequent for the small events,
including C-type flares.  For region-by-region forecasts, if they are not 
assigned, they are not considered as part of the prior flare parameters.
For full-disk forecasts, they are included in evaluation but not in the 
training, leading to a systematic under prediction.

\paragraph{\it Validation:} is performed on demand, producing lists
of standard skill scores (Figure~\ref{fig:daffs_metrics}); Reliability
and ROC plots can be generated automatically, as well ({\it e.g.}
Figure~\ref{fig:daffs_roc}). 

\paragraph{\it Customization:} \daffs\ can be customized for event definition,
timing of forecasts, and forecast validity periods.  Additionally, the categorical 
forecasts can be optimized against either of the two error types (thus minimizing False Alarms
or minimizing Missed Events).

Of note, data which are not retrieved for the near-real-time forecasts, for whatever reason,
are queued for retrieval later to ensure a reasonably complete NRT data source database for
training purposes. 

\subsection{\daffs\ Results}
\label{sec:daffs_results}

\begin{figure}
%\centerline{\includegraphics[width=0.98\textwidth,clip, trim = 0mm 0mm 0mm 0mm, angle=0]{DAFFS_SS_20140701_20170630_2354.eps}}
\centerline{\includegraphics[width=0.98\textwidth,clip, trim = 0mm 0mm 0mm 0mm, angle=0]{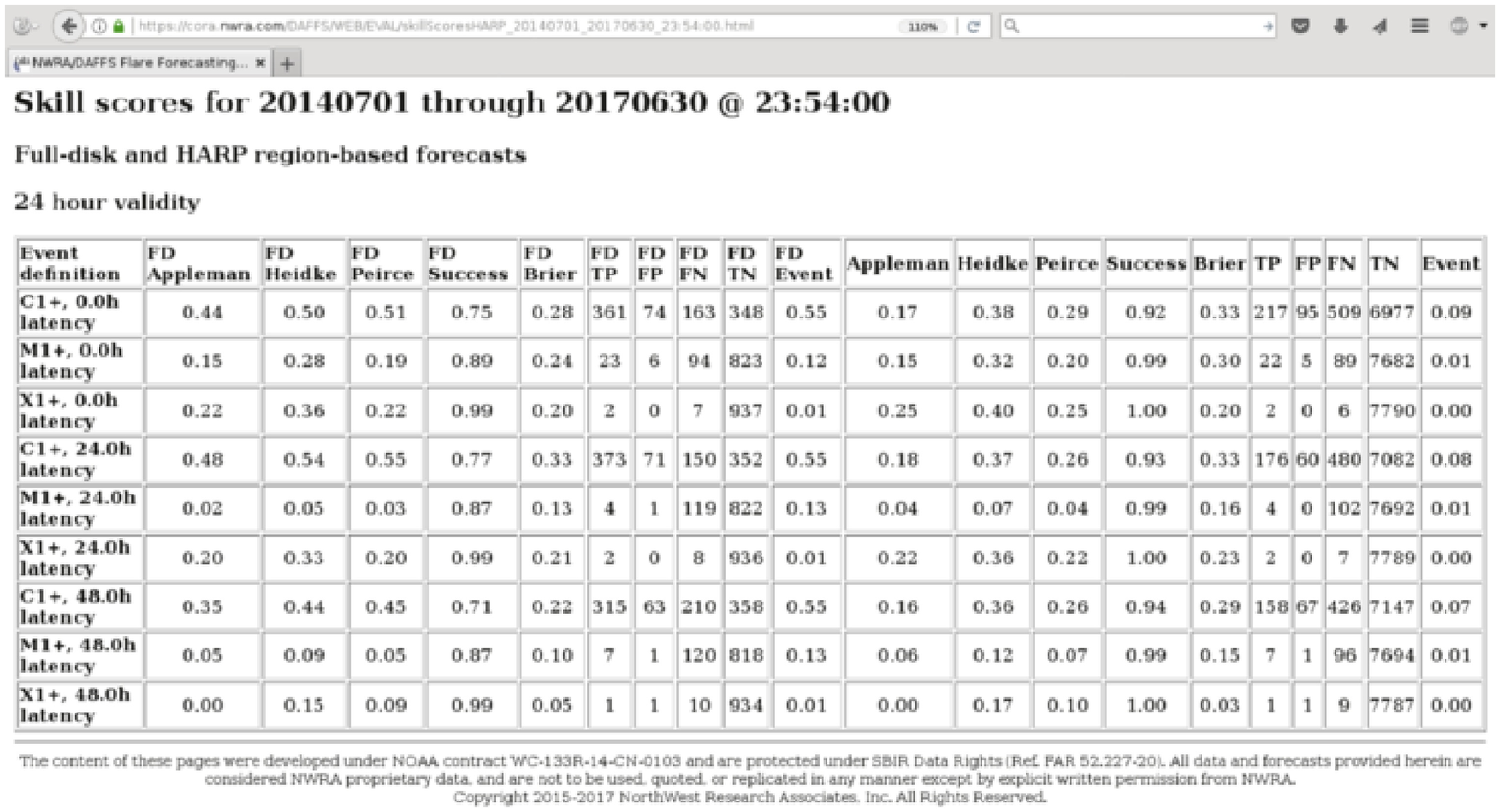}}
\caption{To summarize the metrics for the near-real-time \daffs\ we show a screen-shot from
the output of the \daffs\ self-evaluation code.  Indicated are the evaluation period
(20140701 -- 20170630), for which issuance time (23:54:00 UT), and then 
a variety of skill scores.  Note that the \TSS\ quoted here is {\it not} the ``Optimal''
quoted above, it is evaluated with $P_{\rm th}=0.5$, which is the system default.
``Event'' means ``event rate''; ``FD'' indicates full-disk 
(rather than region-by-region) forecasts.}
\label{fig:daffs_metrics}
\end{figure}

\begin{figure}
\centerline{
\includegraphics[width=0.33\textwidth,clip, trim = 0mm 0mm 0mm 0mm, angle=0]{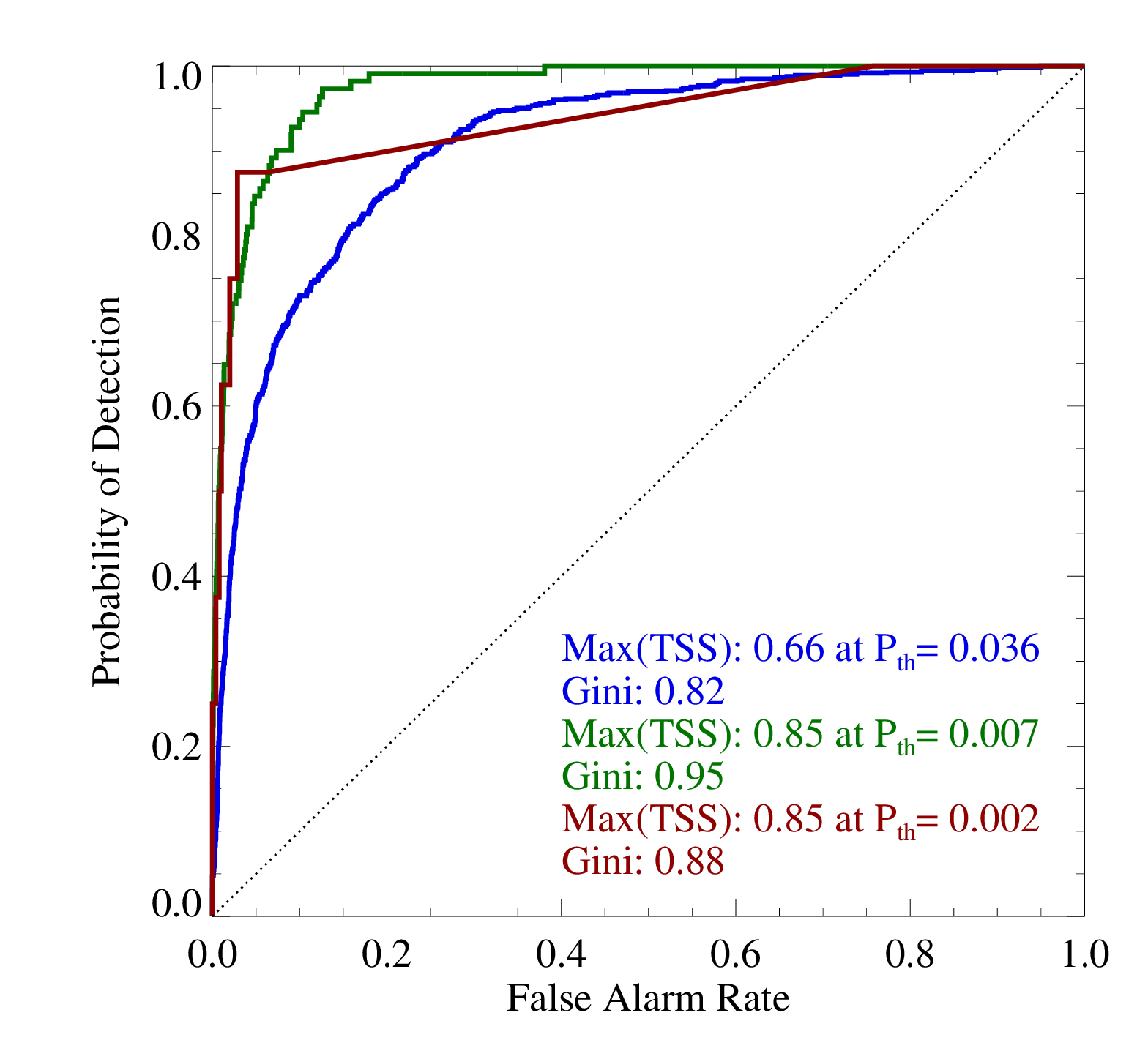}
\includegraphics[width=0.33\textwidth,clip, trim = 0mm 0mm 0mm 0mm, angle=0]{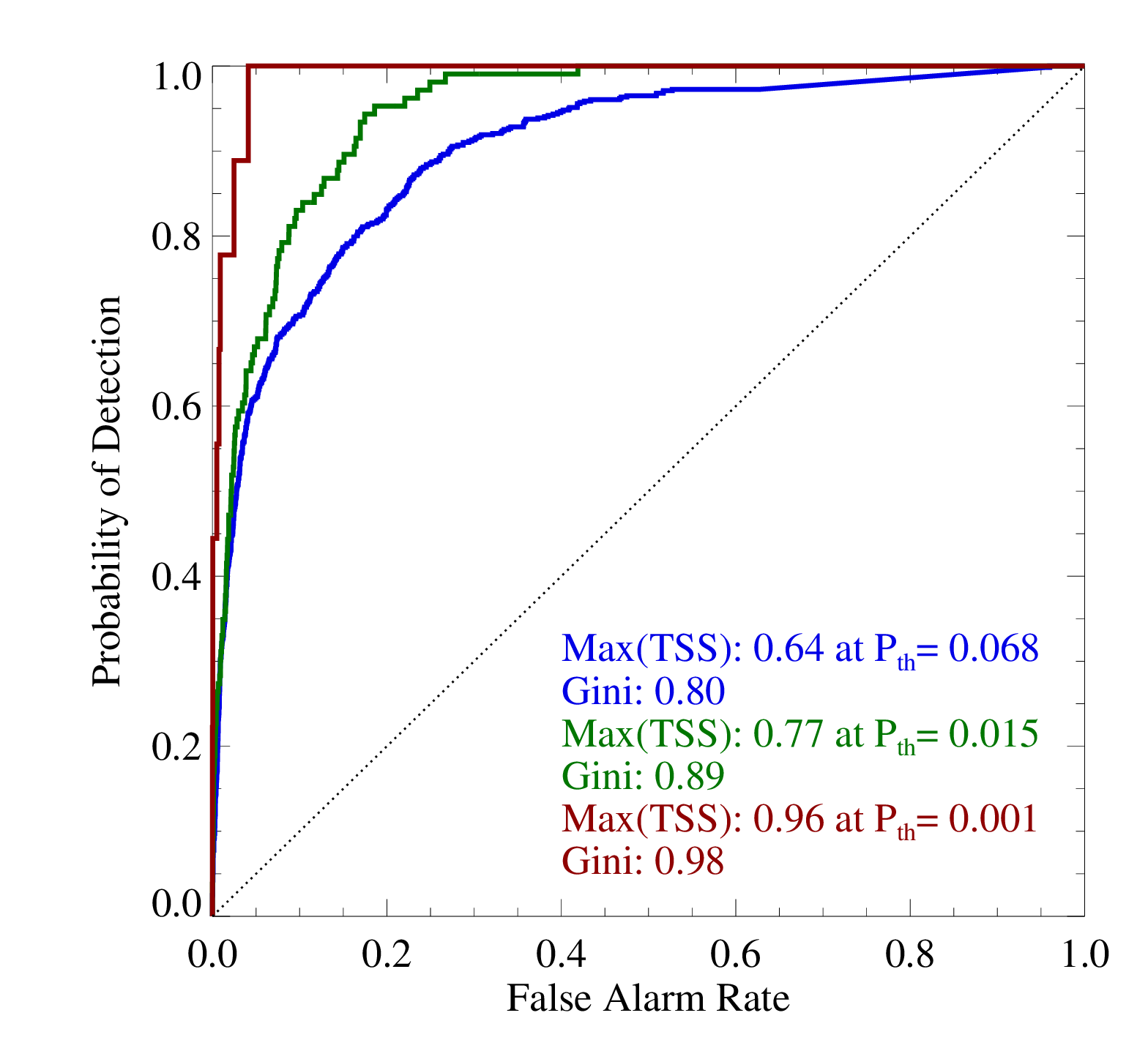}
\includegraphics[width=0.33\textwidth,clip, trim = 0mm 0mm 0mm 0mm, angle=0]{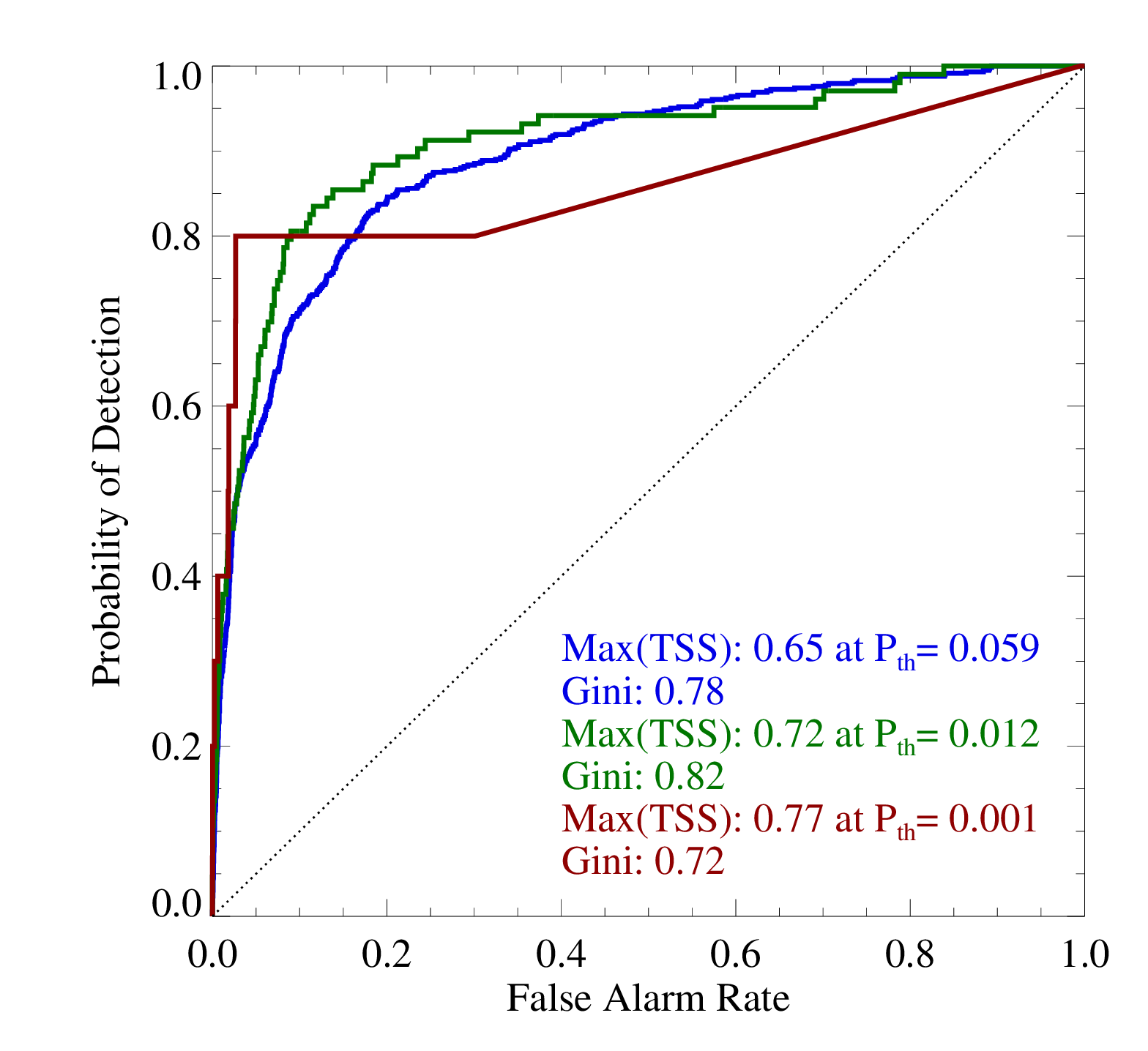}}
\centerline{
\includegraphics[width=0.33\textwidth,clip, trim = 0mm 0mm 0mm 0mm, angle=0]{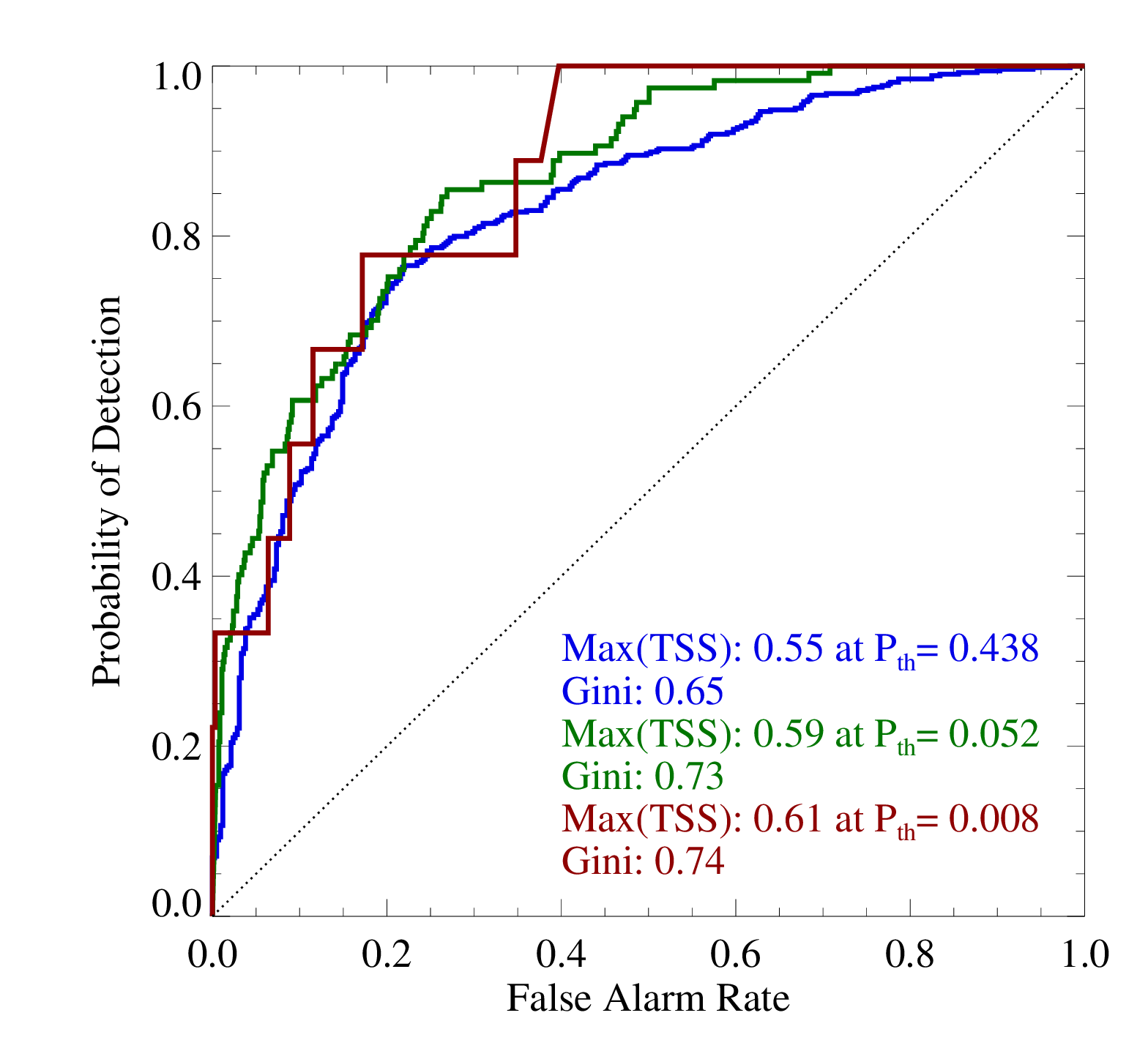}
\includegraphics[width=0.33\textwidth,clip, trim = 0mm 0mm 0mm 0mm, angle=0]{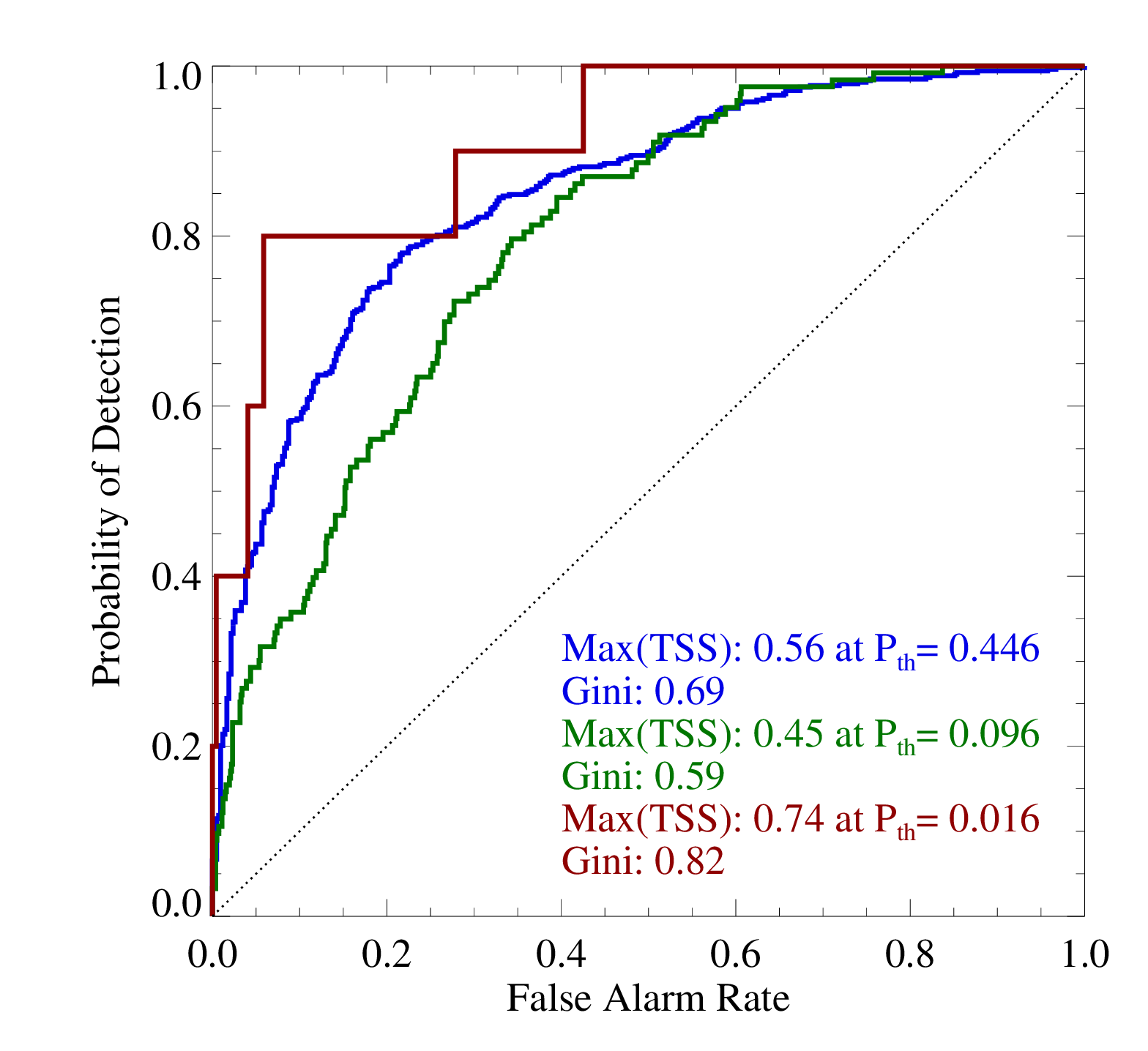}
\includegraphics[width=0.33\textwidth,clip, trim = 0mm 0mm 0mm 0mm, angle=0]{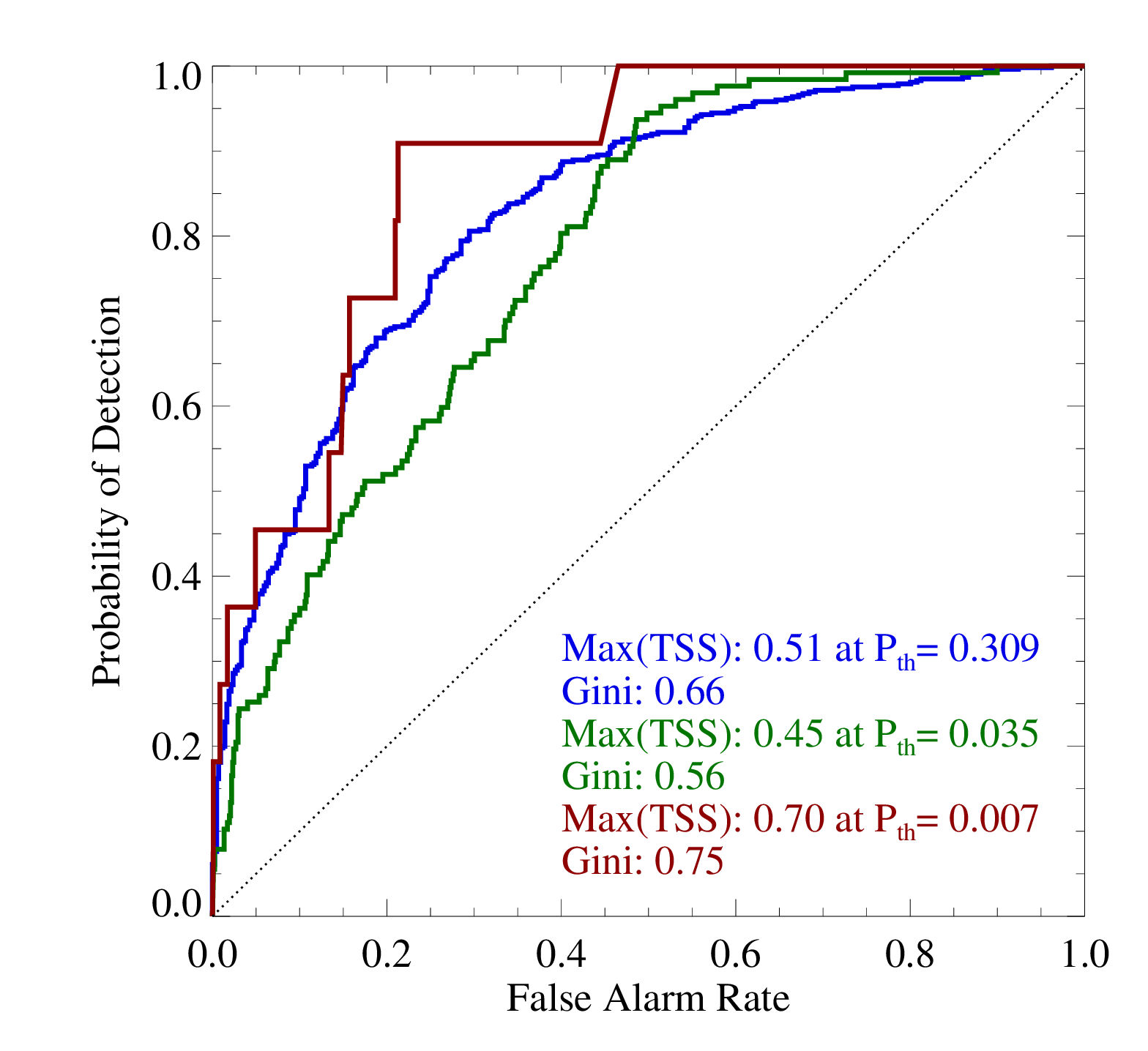}}
\caption{ROC Curve for
\daffs\ near-real-time forecasts (see Figure~\ref{fig:daffs_metrics}).
Blue/Green/Red curves indicate the three event definitions (\CC, \MM,
\XX) for effectively 0\,hr, 24\,hr, and 48\,hr latencies (left:right).
Top row: region-by-region, Bottom row: full disk forecasts.  Also noted
on each plot are the Gini coefficients, the maximum \TSS\ (TSS) achieved, and the probability
threshold $P_{\rm th}$ used for that maximum \TSS\ score.}
\label{fig:daffs_roc}
\end{figure}

For very few of the event definitions do the \ApSS\ scores show
substantial improvement over climatology (Figure~\ref{fig:daffs_metrics}),
although the ROC plots and $G1$ coefficients demonstrate some
performance significantly away from the ``no skill'' ($x=y$) line.
(Note that the Peirce (\TSS) scores quoted in the \daffs\ evaluation,
Figure~\ref{fig:daffs_metrics}, use $P_{\rm th}=0.5$ while the ROC plots
quote the maximum \TSS\ achieved by varying the $P_{\rm th}$.)
Uncertainties are not (yet) quoted when generated from within \daffs,
but the magnitude of the uncertainties in Table~\ref{tbl:nci_region_metrics} can
be a guide.

For probabilistic forecasts, the performances are worse for the larger
events (due to smaller sample sizes). In fact overall, the larger the event
and the longer the latency, {\it generally} the worse the performance,
which is typical \citep{allclear,Murray_etal_2017}.

\subsection{Performance Context}
\label{sec:daffs_perf}

The results above reflect the general performance of the baseline
near-real-time \daffs\ forecast tool.   The results will change
according to the climatology and training interval, and as such should
be interpreted with some care.  These results are also not
directly comparable to numbers quoted for other methods or even to
the \nci\ results above, as discussed in \citet{allclear}, because of
differences in samples, event definitions, and evaluation intervals
(although full-disk forecasts negate a few of these differences).

That being said, \daffs\ is running as an autonomous tool designed
to address operational needs.  During the SBIR Phase-I (feasibility)
study, an \nci-based demonstration out-performed the NOAA/SWPC forecasts
as judged by Brier skill scores that evaluated head-to-head comparisons
with matched event definitions and testing intervals, as required by the
topic description (sec.~\ref{sec:intro}).  This success enabled Phase-II
(prototype development) funding and the near-real-time operational
forecasting tool described here as \daffs.

\section{Future Developments}
\label{sec:future}

The described infrastructure has been designed for flexibility at various 
junctures, and our hope is to implement improvements at a number of them.

Regarding \nci\ in general, we intend to investigate how to optimize the
adaptive-kernel NPDA, where the smoothing parameters are a function of
parameter density.  This should allow a better estimation of the PDF
in high-kurtosis distributions -- a useful additional option for all
scientific questions for which \nci\ might be applied.    Investigating
AKNPDA is a proposed task for future funding.

Regarding \daffs\ specifically, we intend to finish implementing the
GONG-based secondary forecasting -- including then a direct comparison
of performance as compared to the system when HMI vector field data
are available.  A bootstrap approach for the training data will be
implemented to provide estimates of the performance uncertainties.
The event definitions can be modified with respect to forecast
interval, latency, and event limits, and these will be acted upon as 
requested.  As mentioned above, parallel forecasts reporting on multiple
top-performing combinations can be implemented for the NRT forecasting
as a further check against statistical flukes from appearing in the
predicted probabilities.  Additionally, the discriminant analysis
threshold can be optimized according to the costs of either type of
error, or to maximize a particular skill score.  While this is not
widely used when testing the efficacy of new parameters, it can be of
particular importance to customers of the near-real-time tool.  Of note,
these proposed enhancements all have some degree of research-based known value
to add, but will require resources to implement.
Hence, \daffs\ is not open source and is not freely available, as there is no
automatic funding for NWRA (a small business) to continue its support.
However parties interested in using it can be granted limited access
for trial periods, and ``access only with technical support'' levels of
contracts are available for very minimal resource levels.

\nci\ is a research platform by which many questions can be addressed
through the quantitative analysis of appropriate sample sizes.
Forecasting solar flares and energetic events is one such question; it
is fairly well accepted that the present forecasting methods, \daffs\
included, are performing above climatology -- but not performing
particularly {\it well}.  The reasons why this is so are starting to become clear:
likely culprits include a combination of human-defined events and simply
a limited amount of information from photospheric magnetic field data
that themselves may not directly be related to the flare initiation activity
\citep{EEG_chapter}.  As different events, data, and approaches are
investigated we invite further collaborative efforts using \nci\ 
as means to quantitatively test proposed improvements in establishing
distinguishing characteristics of flare-imminent active regions.

\subsection{Other Research Modules}

Described above are the data, parametrizations, and analysis results
for flaring {\it vs.} flare-quiet active regions using Discriminant
Analysis through the \nci.  For other \nci-based research topics,
modules are developed, analyzed, and evaluated in a very similar manner:
parametrizations of the target data are performed and evaluated using
NPDA against relevant event definitions.  The performance can be judged
against that of the modules and results presented here, for example.
Other parameters related to flare productivity have been tested including
helioseismic-derived parameters \citep{ferguson09,Komm_etal_2011a},
plasma-velocity parametrizations \citep{Welsch_etal_2009},
and identification of trigger regions \citep[][based on
\citet{Kusano_etal_2012,Bamba_etal_2013}]{Bamba_etal_2018}.  Parameters
related to the character of the flares could be tested, such as the
ratio of short/long wavelength GOES flux \citep{WinterBala2015}.

Research regarding different event definitions can be constructed within
the \nci\ framework, according to (for example) a CME or lack thereof,
the duration of events and total energy released, whatever appropriate
database of ``events'' is available.  Presently the NOAA-defined event
catalogs form the basis of the GOES event definitions, but this itself
could be modified to use event catalogs based on RHESSI or SDO/EVE events,
for example.

Of note, the \nci\ is not solely useful for flare research.  Topics which
have been investigated within the \nci\ framework include pre-emergence
signatures \citep{trt_emerge3}, a research topic and approach that is ongoing,
and filament eruption \citep{Barnes_etal_2017}.  The \nci\ is not,
we stress, in and of itself a forecasting tool; it uses Discriminant
Analysis to evaluate how well samples from two known populations can be
distinguished.  Indeed, \nci\ may be used with any appropriately defined 
populations for diverse investigation topics.
We actively invite 
collaboration to use the \nci\ framework in broad topics of solar physics.

\section{Summary}
\label{sec:disc}

An investigative infrastructure which has been developed at NorthWest
Research Associates based on Discriminant Analysis classifiers has been
described, and briefly demonstrated in the context of research centered
on distinguishing flare-ready from flare-quiet solar active regions.
The description of the \nci\ and results from the recent flare-related research effort
provide updates to the original publications on this topic.

The Discriminant Analysis Flare Forecasting System \daffs\ is also presented 
here, a near-real-time forecasting tool which germinated from \nci\ and 
related research \citep{dfa,dfa2,dfa3} to address an expressed
need \citep{SWJ}.  As a matter of practicality, \daffs\
by default mimics the system established at NOAA/SWPC in terms of 
event definition and output, although it does not have to.
Many of the details which make this an operational
system (and thus very different from the \nci) are described,
and early results are presented.

\daffs\ is presently in use by the Chief Observers of the {\it Hinode} 
mission.  While the primary data source (SDO/HMI) has an uncertain 
lifetime, \daffs\ was designed to continue without those primary data
although the long-term performance degradation in that situation has yet 
to be determined.  With the oncoming solar minimum and as-of-yet no defined
SDO follow-on mission, \daffs\ will be supported and maintained to the
extent resources allow; it could be of distinct value to new, limited 
field-of-view facilities and missions slated for operation as solar 
activity increases again.

\begin{acknowledgements}
The \nci\ was developed with funds from numerous sources, including
AFOSR contracts F49630-00-C-004 and  % AFOSR 0004
F49620-03-C-0019, % AFOSR 0019
NASA contracts NNH12CG10C, % NASA B619
NNX16AH05G, and % NASA B359
NNH09CE72C, %which is NASA 428
and NSF Grant 1630454. % NSF 399
The \daffs\ tool was developed under NOAA SBIR contracts WC-133R-13-CN-0079 
(Phase-I) and WC-133R-14-CN-0103 (Phase-II) with additional support from 
Lockheed-Martin Space Systems contract \#4103056734 for Solar-B FPP Phase E 
support.  The authors also acknowledge NWRA internal development funds and the efforts
of the two referees to improve the presentation of this work.  
%The editor thanks Brian Welsch 
%and an anonymous referee for their assistance in evaluating this paper.
\end{acknowledgements}

%\bibliographystyle{swsc}
%\bibliography{kdl_biblio}

\end{document}